\documentclass[apj]{emulateapj}

\newcommand{\lumnoh}{\mathrm{erg}\,\mathrm{s}^{-1}}

\newcommand\degrees[1]{\ensuremath{#1^\circ}}

\usepackage{natbib}

\submitted{Accepted Aug 1, 2010}

\begin{document}
\title{Tidal Torquing of Elliptical Galaxies in Cluster Environments}
\shorttitle{Tidal Torquing of Elliptical Galaxies in Clusters}
\author{Maria J. Pereira\altaffilmark{1,2} and Greg L. Bryan\altaffilmark{1}}
\altaffiltext{1}{Columbia University, Department of Astronomy, New York, NY, 10025, USA}
\altaffiltext{2}{Steward Observatory, Tucson, Arizona, 85716, USA}
\email{pereira@astro.columbia.edu}

\begin{abstract}

Observational studies of galaxy isophotal shapes have shown that galaxy orientations are anisotropic: a galaxy's long axis tends to be oriented toward the center of its host. This radial alignment is seen across a wide range of scales, from galaxies in massive clusters to small Milky Way type satellite systems. Recently, this effect has also been detected in dark matter simulations of cosmological structure, but the degree of alignment of dark matter substructures in these studies is significantly stronger than seen in observations. In this paper we attempt to reconcile these two results by performing high-resolution numerical experiments on N-body multi-component models of triaxial galaxies orbiting in an external analytical potential. The large number of particles employed allows us to probe deep into the inner structure of the galaxy: we show that the discrepancy between observed galaxies and simulated dark matter halos is a natural consequence of induced radial shape twisting in the galaxy by the external potential. The degree of twisting depends strongly on the orbital phase and eccentricity of the satellite, and it can, under certain conditions, be significant at radii smaller than the dark matter scale radius. Such internal misalignments will have important consequences, both for the dynamical evolution of the galaxy itself, and for mass modeling of galaxies in clustered environments.

\end{abstract}



\section{Introduction}

In our current picture of the Universe, structure evolves hierarchically from an initial density field that is very nearly homogenous. Tiny perturbations in this field seed the growth of overdensities, which then accrete more matter through gravitational collapse. As matter is pulled into these regions it is continually torqued by the surrounding large scale tidal field; these primordial torques are believed to be responsible for the orientations and angular momentum of galaxies at formation \citep{Peebles:1969p393}. 

Recent work seems to indicate that tidal torques are an important effect at late times as well, acting on galaxies as they fall into larger potential wells and merge hierarchically to form galaxy groups and clusters. These torques should affect a galaxy's orientation, and in fact recent observational studies have identified a significant correlation between a galaxy's position angle and its surrounding tidal field. These group and cluster samples, selected mainly from the Sloan Digital Sky Survey (SDSS) \citep{York:2000p1579}, show a distribution of galaxy orientations that is not isotropic, as had been long assumed; instead, the galaxies' long axes appear to preferentially point toward the center of their hosts \citep{Pereira:2005pL21,Agustsson:2006pL25,Faltenbacher:2007pL71}. 

Subsequent numerical studies have been able to reproduce these results \citep{Kuhlen:2007p1135,Pereira:2008p825,Faltenbacher:2008p146} for dark matter subhalos and to further establish that the anisotropy is caused by a dynamical process occurring inside the host cluster and which is strongly correlated with a subhalo's orbital phase \citep[herafter PBG08]{Pereira:2008p825}. This rules out a purely primordial origin, although some radial alignment is seen even at large distances  ($\approx 3   r_{ vir}$) from the cluster;  halos appear to be weakly aligned with the filaments that feed into cluster nodes.

Under closer inspection, a significant discrepancy between the observational and theoretical results emerges that has not yet been resolved. The alignment tendency in observed galaxies is much weaker than that seen for the halos in cosmological simulations (PBG08). While it is true that observed galaxy shapes can be noisy \cite[see][e.g.]{Siverd:2009p0}, measurement error alone cannot account for such a large difference in alignment. If this discrepancy is confirmed, then one must conclude that the cluster potential is capable of inducing significant misalignments between a galaxy's own dark matter halo and its more centrally concentrated, observable, luminous component. Such internal misalignments could have important consequences for galaxy mass and lens modeling, and this possibility should be studied further. 

More generally, torquing from an external tidal field can affect the internal dynamics of galaxies significantly, and it could be disruptive enough to drive morphological evolution of galaxies in clusters, alongside other gravitational interactions such as harassment and stripping. Moreover, the alignments that this torquing induces between observed galaxy shapes and the surrounding tidal field are a serious issue for weak lensing and cosmic shear studies, and they must be characterized and quantified more precisely before the next generation of cosmic shear surveys comes online. 

We need to study this process in more detail - how and when does the torquing occur? Are all components of the galaxy equally affected? Does the torquing induce radial shape twisting or do galaxies react as rigid bodies? Unfortunately, the low resolution of cosmological simulations makes the detailed dynamics of each individual halo impossible to study, and so we must turn to more idealized numerical experiments on single satellite galaxies. These can be run at higher resolution, and can include a dynamically distinct stellar component embedded in the dark matter halo.

We begin (\S \ref{sec:build}) by describing the observational parameters of a typical galaxy in a typical cluster drawn from the sample of \cite{Pereira:2005pL21} (hereafter PK05), which we then use to build a numerical model of an elliptical cluster galaxy. We describe in detail our methodology (\S \ref{sec:met}), the cluster (\S \ref{sec:clmod}) and galaxy (\S \ref{sec:ell}) models used, the tidal truncation applied (\S \ref{sec:trunc}) and the procedure used to initialize the particles in our simulations (\S \ref{sec:edd}). Halos in cosmological simulations are generally triaxial and we describe in section \ref{sec:triax} how we include this intrinsic triaxiality in our model. We then briefly discuss the issue of shape determination (\S \ref{sec:shape3d}) before presenting our results in sections \ref{sec:dmonly} and \ref{sec:st}. 

As a first approach, we examine the behavior of a single triaxial dark matter halo orbiting an external potential in section \ref{sec:dmonly}, and compare these results with our previous work on subhalos drawn from cosmological simulations in PBG08. We then introduce a stellar component in our models in section \ref{sec:st}. We show that this observable, luminous component is also efficiently torqued by the cluster potential in both circular (\S\ref{sec:stcir}) and eccentric orbits (\S\ref{sec:stecc}), although the resulting dynamics can differ substantially from the more extended dark matter component. The cluster tidal field has a significant impact on both the shape and internal dynamics of the satellite galaxy, and we investigate this in more detail in section \ref{sec:shapevol}. We end by discussing the implications of this work (\S \ref{ch:imp}) before summarizing our main conclusions in section \ref{sec:conc}.

\section{Methods}
\label{sec:build}

\subsection{Methodological Approach and Code}
\label{sec:met}

For the purposes of this work, we have chosen to study the behavior of a single galaxy on different orbits around a single cluster. We build these models to be representative of a typical galaxy in a typical cluster drawn from the sample of  PK05, as described in section 2.2. In order to study the internal dynamics of these galaxies we require a live, self-consistent N-body model of their stellar and dark matter distributions (described in \S 2.4). In contrast, the cluster is modeled as a static, spherically symmetric analytic potential (\S 2.3), which makes the problem tractable numerically and also simplifies the geometry of the system.  We do not expect dynamical friction between the cluster and galaxy to be important within these time scales and mass ratios, and the fact that the satellites remain on coplanar, energy conserving orbits makes the dynamics of the system simpler to analyse. 

Once the initial conditions for the stellar and dark matter particles have been found (\S \ref{sec:edd}) we use {\tt GADGET 2.0} \citep{Springel:2005p1105} to evolve the models numerically. We have modified the code in order to adiabatically squeeze the live particles to form a triaxial distribution. Once the desired 3D shape is attained, we turn off the artificial squeezing and let the model evolve in isolation for 1 Gyr, at which point the shape has become stable. We then modify  {\tt GADGET} again in order to impose an analytic potential representing the cluster tidal field, adding an extra acceleration component to each particle at each time step. The galaxy model is placed at the apocenter of the orbit being simulated, and the particle orbits are then numerically integrated for $\sim 15$ Gyr, or 3-5 cluster orbits. 

Our galaxy particles are initialized in isolation. If a strong external field is applied suddenly, the change in the potential will shock the particle distribution and lead to unphysical behavior. We therefore further modify \texttt{GADGET} to allow for the gradual ``turn on" of the cluster potential, mimicking (crudely) the infall of galaxies from the field. The cluster mass is grown over a period of 1 Gyr, and, in order to maintain the halo in the correct orbit through the process, we subtract at each time-step the acceleration induced by the presence of the cluster potential on the halo's center of mass from each particle. Our results are generally insensitive to the details of this initialization: the overall alignment behavior of the halo is reproduced with or without a slow turn-on of the external potential.

\subsection{A Typical Galaxy in a Typical Cluster}
 \label{sec:typ}
The observational sample described in PK05 spans a wide range of cluster and galaxy masses. All analyses to date have shown that the degree of radial alignment does not depend significantly on individual galaxy or cluster properties \citep{Pereira:2008p825,Knebe:2008p517}, and we therefore choose to build a single model representing a typical galaxy orbiting a typical cluster in that sample, where we interpret  typical to mean an object with parameters corresponding to the median of the entire population.

The median redshift of the cluster sample is $z=0.1$ and the median X-ray luminosity $L_X=3.47\times10^{44}$ ergs$/$s (0.1 - 2.4 keV). What does the typical elliptical galaxy in these clusters look like? The distribution of R band absolute magnitudes for the galaxies in these typical clusters has a median $M_{ R} = -20.3$. Their elliptical surface brightness profiles are generally well fit by a de Vaucouleurs law over a wide range of radii (see, e.g., \citet{Burkert:1993p23}): 

\begin{equation}
\log \left ( \frac{ I(R)}{ I(R_e)} \right )= -3.331 \left[ \left(\frac{ R}{ R_e}\right)^{1/4} -1 \right ]
\end{equation}
where $R_{ e}$ is the effective radius, and is roughly equivalent to the half-light radius of the galaxy. The median effective radius for our sample galaxies as fit by the SDSS data analysis pipeline is $R_{ e} = 3.25$ kpc.

\begin{figure*}
\begin{center}
\includegraphics[width=0.85\textwidth]{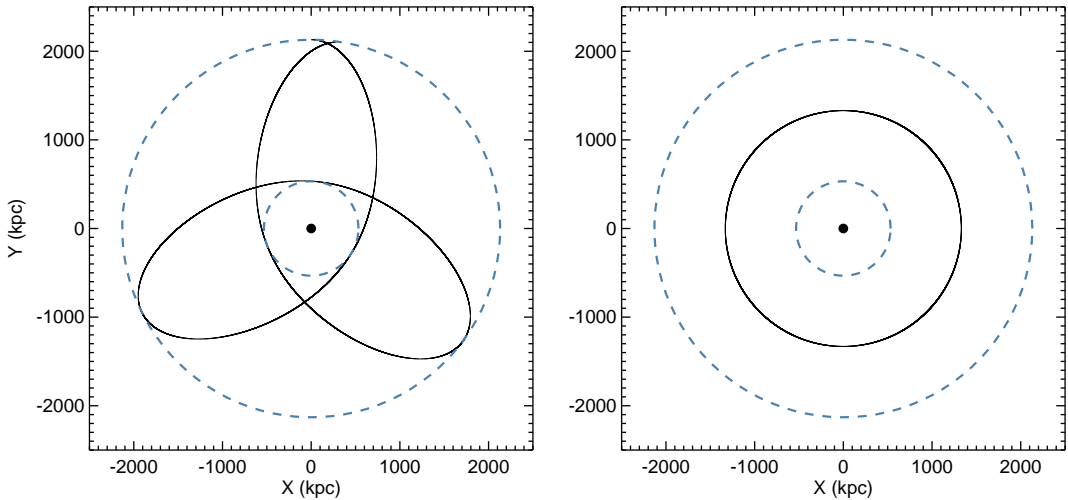}
\vspace{-0.4cm}
\end{center}
\caption[Characteristic Orbits]{\label{fig:orbs1}
Left: The 4:1 eccentric orbit. Right: The circular orbit. Dashed blue circles represent $r_{200}$ (outer) and $r_s$ (inner) for the halo model.}
\end{figure*}

\subsection{Cluster model}
\label{sec:clmod}

We model the host cluster as a static external potential in which our galaxies orbit. We assume an NFW profile for the mass distribution of the cluster, and infer the mass and concentration of the profile from observable parameters of the cluster sample.

There are many observational studies aiming to directly relate X-ray luminosity to total mass for group and cluster sized halos. We use the scaling relation of \cite{Rykoff:2008pL28} who fit ROSAT X-ray luminosities  and weak lensing masses for the MaxBCG catalog \citep{Koester:2007p239} with the following parametrization:

\begin{equation}
 L_X= 12.6^{+1.4}_{-1.3} \left ( \frac{M_{200}}{10^{14}h^{-1}M_{\sun}} \right)^{1.65 \pm 0.13} 10^{42}\,\lumnoh.
\label{eq:LMresult}
\end{equation}

This relation was calibrated for $z=0.25$, and will evolve slightly with redshift. For a fixed mass, we assume the evolution of X-ray luminosity with redshift is passive and follows $L_{ X}(z) \propto \rho_c(z)^{7/6}$, where $\rho_c(z)$ is the critical density of the universe at redshift $z$ \citep{Kaiser:1986p323}. This results in a virial mass at $z=0.1$ of $M_{200} = 8.3.10^{14} h^{-1}  M_{\sun}$ for the median $L_X$ of our sample.

The concentrations of dark matter halos appear to be correlated with halo mass in both simulations and observational studies. \cite{Neto:2007p1450} measured this correlation for a large sample of $z=0$ halos drawn from the Millenium simulation \citep{Springel:2005p629}. The sample spans three decades in mass and is best fit by:

\begin{equation}
c_{200} = 5.26 \left (\frac{ M_{200}}{10^{14}h^{-1}M_{\sun}}\right)^{-0.1}
\label{eq:cvsm}
\end{equation}

Halo concentrations are also observed to decrease with redshift, roughly along $c(z)=c_0/(1+z)$, where $c_0$ is the concentration at $z=0$. Both observed and simulated relations show significant scatter at any particular mass or redshift value, a consequence of the diverse accretion histories of cosmological halos. We adjust this $z=0$ mass-concentration relation to the median redshift of our sample and estimate a concentration for our typical cluster of $c \approx 4$.
The parameters for the cluster NFW potential are then: $M_{200} = 8.3.10^{14} h^{-1}  M_{\sun}$, $c = 4.0, r_s = 532.8$ kpc and $r_{200}= 2131.1$ kpc.

\subsection{Characteristic Orbits}

According to cosmological simulations of large scale structure, most galaxies are fed into clusters through filaments. This leads to a radial distribution of orbits in the outskirts of clusters, which progressively becomes more isotropic closer to the cluster center. Nevertheless, even for a perfectly isotropic orbital distribution the average eccentricity of these orbits is high, with typical apocenter to pericenter ratios of 4:1 \citep{van:1999p50}. Dynamical friction progressively shrinks the orbits, but with little circularization - the overall measured eccentricities remain constant. This picture is consistent with most observational studies, which find that galaxy orbits in clusters appear isotropic \citep{Hwang:2008p218}, although there have been claims (see, e.g. \cite{Biviano:2009p0}) that different types of galaxies appear to have orbits drawn from different distributions, with early-types appearing generally isotropic, and late-types biased towards a more radial distribution. 

The simulated clusters presented in PBG08 also show large orbital eccentricities in their substructure, with an average apocenter to pericenter ratio of 4:1, so we take this as the eccentricity  of our representative orbit, with $r_{\rm apo} =  r_{ 200, \rm host} = 2131.1$ kpc and $r_{ \rm peri} = r_{ 200, \rm host}/4.0 = 532.8$ kpc .

As a first approach, we will simplify the system even further and place the galaxy on a circular orbit, in order to first understand how our galaxy reacts to an external tide of constant magnitude. The radius of the  circular orbit was chosen so that the orbit has the same energy as the representative eccentric orbit, i.e. $r_{\rm cir} = 5  r_{\rm peri} /2 = 1332.0$ kpc, and $t_{\rm orb} \approx 5.2$ Gyr (see fig \ref{fig:orbs1}).

\subsection{Elliptical Galaxy Model}
\label{sec:ell}

We want to create a physical model of a galaxy which has an absolute magnitude of $M_{r} \approx -20.3$ and is well fit by a de Vaucouleurs profile with  $R_{ e} \approx 3.25$ kpc. 

The galaxy luminosity in the r band is then: $L_{\rm gal,r} = L_{\sun ,\rm r} \times 10^{\frac{ M_{\sun  ,\rm r}-  M_{\rm gal, r}}{2.5}} = 1.1\times10^{10} L_{\sun , r}$, with $M_{\sun , r} = 4.76$ \citep{Blanton:2003p2348}.  According to the study by \cite{Kauffmann:2003p33} mass-to-light ratios vary steeply with absolute magnitude for galaxies observed with SDSS filters. For a galaxy with $M_{ r} \approx -20.3, (M/L)_{\rm r} =3.16$, which gives a stellar mass for our galaxy of $M_{\rm gal} \approx 3.3\times 10^{10} M_{\sun}$.

The de Vaucouleurs profile has no analytical 3D deprojection, but there are a few two-power density profiles that give projected light distributions that follow the r$^{1/4}$ law at most radii. We pick one of the most commonly used, the Hernquist profile  \citep{Hernquist:1990p359}, since it has a simple analytical form both in density and potential and is finite in mass:

\begin{equation}\label{eq:bulgeprofile}
{\rho}_{ stars} =  \frac{ M}{2\pi a}
\frac{1 }{\left (r/a\right )
\left (1 + r/a\right )^{3}}~.
\end{equation}

In this equation, $a$ is the scale length, and it is related to the projected effective radius, $R_{ e}$ by $R_{ e} = a (1+ \sqrt 2)$. Combining this with our estimate of the total stellar mass completely defines the stellar density profile. 

We model the dark halo with the more theoretically motivated NFW profile \citep{Navarro:1997p493}:

\begin{equation}\label{eq:haloprofile}
{\rho}_{ halo} = 
\frac{\delta_c \rho_c }{\left (r/r_s\right )
\left (1 + r/r_s\right )^{2}}~.
\end{equation}
Here, $\rho_c$ is the cosmological critical density, $\delta_c$ is a characteristic dimensionless density and $r_s$ is the scale radius. The NFW mass profile diverges logarithmically with $r$, which means that: a) in practice the halo must be truncated at a large radius (this procedure will be described in \S \ref{sec:trunc}), and b), the mass of the halo must always be defined within a specific radius. 

There are a few different masses commonly quoted. Here, we define the mass of the halo as $M_{200} = 200\times \rho_c \frac{4}{3} \pi r_{200}^3$ where $\rho_c$ is the cosmological critical density, and $r_{200} = c  r_s$, where $c$ is the concentration of the dark matter halo. These definitions imply:

\begin{equation}\label{eq:delta_c}
{\delta}_c =\frac{200}{2}\frac{c^3}{[\ln (1+c)-c/(1+c)]} 
\end{equation}
which means that, for a given $M_{200}$, there is a single free parameter in the NFW profile, which can be expressed in terms of $\delta_c$ or $c$.

Of course, the precise shape and total mass of our galaxy's dark component is not well constrained by our observations. We make a few simplifying assumptions: We require the circular velocity profile of the galaxy to be approximately flat over the observable range $r \sim 0.5-3.0 R_{ e}$, in order to agree with the dynamical models of elliptical galaxies of, e.g. \cite{Gerhard:2001p1936}, and we use the mass-concentration relation of equation \ref{eq:cvsm} to calculate $c$.

These constraints limit the total mass of the halo to a narrow range, $M_{200} \sim 6\times 10^{11} M_{\sun}$. From  the mass all other parameters follow: Our fiducial model has $M_{200} = 6\times 10^{11} M_{\sun}, c = 7.9,  r_s = 22.6$ kpc and $r_{200}= 178$ kpc.

\subsection{Tidal Truncation}

\label{sec:trunc}
The NFW profile is nothing more than a fit to halo density over a certain radial range. At large radii, the profile becomes unphysical, with the mass diverging logarithmically. It therefore follows that even isolated N-body halos, when drawn from an NFW density profile must be truncated at some arbitrary large radius. 
The presence of the external tidal field in our model sets a natural scale for truncation, since it will strip away most material which lies at radii larger than the tidal radius, r$_t$.  We approximate r$_t$ as the Roche radius of a spherical system under the distant tide approximation:
\begin{equation}
 r_t = \left (\frac{m}{3M}\right)^{1/3} R_0,
 \end{equation}
where $R_0$ is the distance to the cluster center, $m$ is the mass of the satellite and $M$ the mass of the host enclosed at radius $R_0$. 

The exact form of the truncation is unimportant for our purposes, but it cannot be too sharp. Simply setting $\rho =0$ at $r > r_t$ would lead to unphysical behaviour in the satellite model. Instead we apply a smooth cut-off to the radial density profile:

\begin{equation}\label{eq:haloprofiletr}
{\rho}_{ halo} = 
\frac{\delta_c \rho_c }{\left (r/r_s\right )
\left (1 + r/r_s\right )^{2}\cosh(r/r_{\rm trunc})} ~.
\end{equation}

Galaxies on different orbits will suffer different degrees of stripping. We choose our truncation radius so that, for the galaxy on the circular orbit, $80\%$ of the galaxy's initial mass lies within $r_{\rm t} (r_{\rm cir}) \approx 80$ kpc. The same galaxy at apocenter on an eccentric orbit will have  $95\%$ of its mass enclosed within $r_{\rm t}(r_{\rm apo})$. These requirements set $r_{\rm trunc} = 2.2 r_s$.

\subsection{Particle Initial Conditions: Eddington Inversion }
\label{sec:edd}

Once the density profile of stars and dark matter is specified, particle positions can be drawn to match these radial profiles using sampling techniques. However, a full set of initial conditions for N-body simulations requires the velocities of each particle to also be specified, and these must be such that the system is already in equilibrium. 

In the case of a particle system that is described by a known, spherically symmetric potential, $\Phi(r)$, a distribution function, $f$,  can be constructed that depends solely on the two integrals of motion: the binding energy, $E$,  and the angular momentum vector, ${\bf L}$, per unit mass. For a non-rotating, isotropic system (which we assume here), ${\bf L} =0$, and $f=f(E)$. The distribution function can then be calculated using Eddington's formula \citep{Eddington:1916p572}:

\begin{equation}
f(E) =\frac{1}{\sqrt{8} \pi^2}\left[ \int_{0}^{E}\frac{{ d}^2 \rho}{{ d} \psi^2} \frac{{ d}\psi}{\sqrt{E-\psi}} + \frac{1}{\sqrt{E}} \left (\frac{{ d} \rho}{{ d}\psi}\right)_{\psi=0}\right] \ ,
\label{E_df}
\end{equation}
where $\rho(r)$ is the density profile, and 
$\psi(r)=-\Phi(r)$ is the relative gravitational potential.
The second term of the right-hand side vanishes for any sensible behaviour of $\psi(r)$ and $\rho(r)$ at large $r$, and can be ignored.

This formula can also be used in the case of two spherical distributions which are in equilibrium with each other, as is the case in our galaxy model. Then, the potential, $\Psi$, should be replaced by the combined potential of the two components $\Psi_{\rm TOT} = \Psi_{\rm HALO} +\Psi_{\rm STARS}$, and the integral is then solved separately for each radial density profile. The distribution functions obtained will be for two particle distributions that are in equilibrium. 
The procedure to calculate $f(E)$ involves first calculating the cumulative mass function for each component, then using these to calculate the total  combined potential for the galaxy. This is gridded at regular intervals in $\log r$ and spline interpolation is used to obtain $\Psi_{\rm TOT}(r)$. This is then numerically integrated using an adaptive integration routine (GSL: Galassi et al. 2006), and the values of $f(E)$ are tabulated at regular intervals in log$(E)$. 

Once $f(E)$ has been determined for both components, it is straightforward to populate the galaxy with particles using an acceptance-rejection algorithm (see, e.g. Press et al. 1996). Briefly, we start by drawing random values for R from a uniform distribution $R \in [0,R_{\rm max} =2 R_{200}]$ and calculating $\Phi(R)$. Random values of $V$ are then drawn from a uniform distribution $V \in [0,V_{max}=\sqrt{-2\Phi(0)}]$ and the binding energy, $E$, of the particle is calculated. If $E>0$, the particle is unbound and rejected. Otherwise, the normalized $f(E)$ is calculated and compared to a random value, $x$, drawn from a uniform distribution $x\in [0,1]$. If it is smaller than $x$, the particle is kept, if larger, the particle is rejected, and the process starts over again, until the galaxy has been populated.

The halo is populated with particles of mass $m_H =  6\times 10^5 M_{\sun}$, for a total of $N_H=545420$ particles, giving a total, truncated mass of $M_H = 3.2\times 10^{11} M_{\sun}$. The stellar particles have masses $m_S = 1.5\times 10^5 M_{\sun}$. There is a total of $N_S =220000$ stellar particles for a total mass of $M_S = 3.3\times 10^{10} M_{\sun}$. This is enough particles so that two-body relaxation effects are unimportant for $r \ge1$kpc  for a period $T> 20$ Gyrs, which is approximately the total amount of integration time required for each of our experiments. We check the stability of our models by evolving them numerically in isolation for 20 Gyrs and verifying that their shape and density profile does not evolve significantly.

We use a separate particle softening length for the halo and bulge. For the halo, $r_{\rm soft} = 0.1$ kpc $\approx \frac{r_{200}}{\sqrt{N_H}}$ (see, e.g. \citet{Power:2003p14} for a discussion of optimal softening length). Then, following \cite{McMillan:2007p541}, we pick a softening length for the stellar particles $r_{\rm soft} = 0.05$ which ensures that the maximum force exerted by every particle ( $\propto m_i/r_{\rm soft}^2$) is the same. 

\subsection{Triaxial Squeezing}
\label{sec:triax}

Triaxial halos are seen in most cosmological simulations and are believed to be a natural consequence of gravitational collapse and hierarchical merging. Unfortunately, finding conclusive observational evidence for triaxial dark matter shapes will be incredibly difficult: the most promising current studies, which use tidal streams as particle tracers of the Milky Way potential, are still in their infancy. On the other hand, there are now many studies, both photometric and kinematic that seem to indicate that elliptical galaxies have shapes that are intrinsically triaxial \citep{Lambas:1992p404,Franx:1991p112}.

There is no simple recipe that can make a triaxial NFW N-body halo, since there are no known analytical distribution functions for systems that are both cuspy and triaxial. Some commonly used methods involve collapsing a self-gravitating diffuse collection of particles numerically (e.g. \cite{Norman:1985p20}) or colliding two cuspy spherical N-body halos to create a single cuspy triaxial one (e.g. \cite{Moore:2004p522}) but all these have the significant drawback of not being able to control the shape of the final galaxy with any precision. An alternate method, known as adiabatic squeezing, was proposed by \cite{Holley-Bockelmann:2001p862} and refined later by \cite{Widrow:2008p1232}. The main advantages of this method is that not only is it applicable to any equilibrium spherical distribution, one has reasonable control over the final triaxiality of the system. We apply it here to our spherical model.

\begin{figure*}
\begin{center}
{\includegraphics{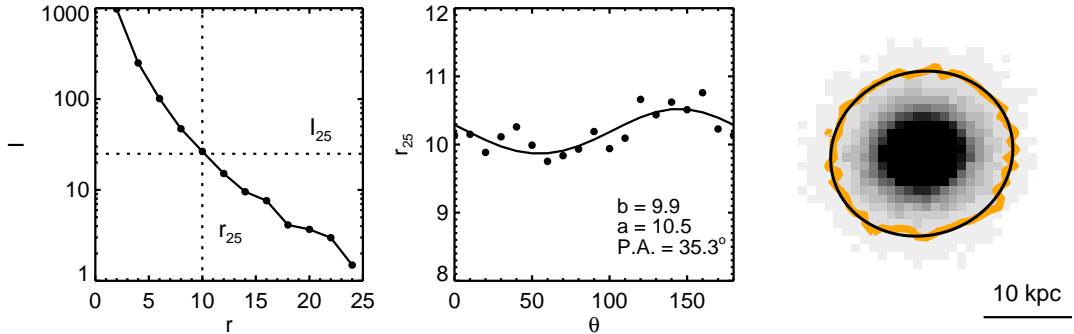}}
\end{center}
\caption[Isophotal Fitting Procedure]{\label{fig:isophot}\footnotesize{Isophotal fitting procedure for a typical galaxy snapshot. From Left to Right: a) Projected surface brightness profile along a specific azimuth. Linear interpolation is used to calculate $r_{25}$. b) Elliptical fit to $r_{25}$ vs azimuthal angle. c) Elliptical parametrization is generally a good fit to the isophotal contour (in orange).}}
\end{figure*}

\subsubsection{Description of Method}

Following \cite{Widrow:2008p1232} we numerically squash our spherical particle distribution to the desired triaxial shape. This is done adiabatically and respecting energy conservation which ensures that the system's differential energy distribution is preserved. The phase-space distribution function is numerically altered, but the radially averaged density and velocity profiles remain stable throughout the process. We modify \texttt{GADGET} to incorporate the numerical squeezing of the live galaxy. 

In detail, at each timestep, an added acceleration is given to each particle which is perpendicular to its velocity, and obeys the following relations:

\begin{eqnarray}
 a_x& =& \frac{\left (\beta_1(t)-\beta_2(t)\right )v_y^2 + \beta_1(t) v_z^2}{v^2}\,v_x~,\nonumber \\
a_y &= &\frac{\left (\beta_2(t)-\beta_1(t)\right )v_x^2 + \beta_2(t) v_z^2}{v^2}\,v_y~,\nonumber \\
a_z& =& -\frac{\beta_1(t)v_x^2 + \beta_2(t) v_y^2}{v^2}\,v_z
\label{eq:az}
\end{eqnarray}
where $v$ is the speed of the particle and ${\bf a}\cdot {\bf v} = 0$,
as required.  

$\beta_{1}$ and $\beta_2$ determine the final shape of the halo. Roughly, a positive $\beta_{1}$ increases the $x$ axis length, whereas a positive $\beta_{2}$ increases the $y$ axis length. Most relevant for a triaxial system,  $\beta_{1} -  \beta_{2}$ sets the difference between the minor to intermediate axis. In order not to shock the system with a sudden``turn on'' at $t=0$, the $\beta$s increase from zero to their
maximum values over a period $T_G$ with a time-dependence given by
\begin{equation}\label{eq:betatimedependence}
  \beta_i = 
\beta_{i,max}(t)\left (3\left (t/T_G\right )^2
    - 2\left (t/T_G\right )^3\right )~.
\end{equation}
$\beta_1$ and $\beta_2$ then remain constant for a time $T_C$ before
decreasing to zero over a time $T_D$ following the same relation.

$T_G + T_C + T_D$ must be long enough so that the process is adiabatic. In our case we pick $T_G = T_D = 0.5$ Gyr, $T_C=2.5$ Gyr, $\beta_1 = 0.48$ and $\beta_2 = 0.02$, and we evolve the galaxy in isolation for another 15 Gyrs, to make sure that the shape is stable over the timescale relevant to our experiments. 

The minor axis length does not evolve significantly once the squeezing process is completed, staying fairly constant over the timescale of the simulations, apart from a small tendency for circularization that is due to particle noise in the simulation. The intermediate axis, on the other hand, particularly for the dark matter component, shows significant evolution as soon as the squeezing is turned off, coming to a stable state a couple of Gyrs later. It then evolves slightly towards larger values, in the same manner as the minor axis.

\vspace{0.5cm}
\subsubsection{Picking Axis Ratios}

Now that we have the ability to create self-consistent triaxial shapes, we must pick appropriate axis ratios for our model.  We choose to squash stellar and luminous particles simultaneously, so the ellipticity of the halo will be related to the ellipticity of the stellar component. 

Elliptical galaxy shapes have a wide distribution of projected axis ratios. \cite{Lambas:1992p404}  showed that a combination of axisymmetric (purely prolate or purely oblate) systems alone cannot explain this distribution, but that intrinsic triaxial shapes are required. Their best estimate for the distribution from which these triaxial shapes are drawn is a two dimensional Gaussian with $b_0 = 0.95, c_0 = 0.55, \sigma_b=0.35$ and $\sigma_c =0.2$. This corresponds to mean galaxy shapes of $\langle b/a \rangle \approx 0.8$ and $\langle c/a\rangle \approx 0.2-0.7$ within a few times $R_{ e}$.

On the other hand, we can place constraints on the triaxiality of dark matter halos from analyses performed of large cosmological simulations. In particular, a careful analysis of the shapes of substructure galaxies was performed by \cite{Kuhlen:2007p1135}. They observe substructure halos to be less triaxial than their isolated counterparts, with mean axis ratios of $\langle b/a \rangle \approx 0.83$ and $\langle c/a\rangle \approx 0.68$ within $r_{v_{\rm max}} \approx 2.163\times r_s \approx 50$ kpc for our model.

Our fiducial model has stellar axis ratios of $\langle b/a \rangle \approx 0.82$ and $\langle c/a\rangle \approx 0.57$, halo axis ratios of  $\langle b/a \rangle \approx 0.80$ and $\langle c/a\rangle \approx 0.55$ and is therefore a good representation of a typical galaxy according to these constraints.

\subsection{Shape Measurements}

\label{sec:shape3d}

We measure the 3D shape of our galaxy by calculating the \emph{reduced} inertia tensor of the particle distribution:
\begin{equation}\label{eq:inert}
\tilde{I}_{jk} = \sum_i m_i \frac{ r_{i,j} r_{i,k} }{r_i^2}.
\end{equation}
which weights particles equally regardless of their distance to the center of the halo. This is then diagonalized to find the principal axes of the halo. The eigenvectors and eigenvalues of this reduced form of the inertia tensor give us the principal axes of the halo and a measure of their relative lengths ($b/a, c/a$). We follow \cite{Dubinski:1991p496} and introduce two modifications to the algorithm. First, the axes are found iteratively: at each step the halo is rotated into the principal axis frame, the particle distribution and the inertia tensor recalculated. Secondly, instead of a spherical cut-off, an elliptical cropping is used proportional to the eigenvalues calculated in the previous iteration. This returns eigenvalues that are a bit smaller, since it avoids biasing the measurement towards spherical values by imposing an artificial spherical cut-off, and the procedure generally converges rather quickly.

Once stellar particles are included in our model, we can start making more meaningful comparisons to the observed data of PK05. The position angles measured in that study were output by the SDSS pipeline and are the product of model independent isophotal fitting: For each galaxy image, an ellipse is fit to the isophotal contour corresponding to $I_{25} = 25$ mag/arcsec$^{2}$ in the SDSS r band. The position angle of the galaxy then corresponds to the major axis of this ellipse.  Most undisturbed galaxies have isophotes that are very nearly elliptical (with small deviations for boxy and disky galaxy types) and their contours are smoothly continuous so that their position angles and shapes do not vary significantly with radius. For these galaxies the radius of the isophote is relatively unimportant. For galaxies with significant isophotal twisting, this becomes much more crucial, since this method only captures the shape of the galaxy at the radius which it is measured. This is an important difference with respect to the method we have been using until now, in which we calculate the reduced inertia tensor of the entire particle distribution within a certain radius and thus measure a mean shape for the entire halo, unbiased by any particular distance. 

In order to measure isophotal shapes for our simulated galaxy, we first project the stellar particles along the line of sight and use the mass-to-light ratio discussed in \S \ref{sec:ell} to calculate the surface brightness profile of each galaxy along 36 different azimuthal directions at \degrees 5 intervals. We use linear interpolation to find the radial distance at $I_{25}$ for each azimuthal slice, and then fit an elliptical profile to that contour (see figure \ref{fig:isophot}). We find that our squeezed stellar distribution is generally well fit by ellipsoidal contours, with low residuals in the fit. 

It is important to note that our numerical experiments employ a static spherical potential, and that therefore all our orbits are coplanar, and the geometry of the problem is considerably simplified.  In a real cluster, or a simulated cluster generated from the collapse of cosmological initial conditions, hosts are triaxial and dynamically evolving. Orbits precess and decay through dynamical friction, while also being  observed from random angles, so the relevant radial angle is measured in 3d. With the experiments in this paper we are attempting to understand the dynamics of these substructures by watching them evolve in controlled and simplified set-ups. All simulations were initialized so that the major and intermediate axis of the galaxy are within the orbital plane, and the minor axis is normal to it. This is the most stable configuration, and it is maintained throughout the orbit. If the galaxy is initialized such that its minor axis is in the plane of the orbit, while its intermediate axis points out of the plane, these axes will quickly tend to flip, thereby achieving a more stable configuration (Christine Simpson \& Kathryn Johnston, in prep).

\begin{figure}
\begin{center}
{\includegraphics[width=8.5cm]{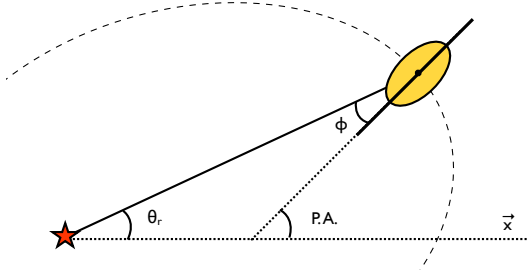}}
\caption[Sketch of angle definitions to be used in the text]{\label{fig:angles}\footnotesize{Sketch of angle definitions used in the text. The star represents the position of the cluster center. [Orbit not to scale!] }}
\end{center}
\end{figure}

In order to quantify the degree of radial alignment, we calculate $\phi$, the angle between the major axis of the galaxy (as determined by one of the 2 methods outlined above) and the direction to the cluster center. We also define the position angle (P.A.) of the galaxy as the angle between its major axis and the fixed \textbf{x} axis in the cluster frame, and $\theta_r$ as the angle between the line connecting the centers of the cluster and satellite and the same fixed axis \textbf{x}, so that $\phi=$P. A.$ - \theta_r$ (figure \ref{fig:angles}).  These angles are all always confined to the orbital plane, by design, so that the problem becomes essentially two dimensional.  Our analysis will not directly address the issue of projection in real cluster observations, where each orbital plane is projected onto the plane of the sky, because we are mostly concerned with the details of the physical mechanism. We note that, while this will dilute any attempt to measure radial alignment in real galaxies, it cannot, on its own, account for the discrepancy between the observations of PK05 and the simulations of PBG08, as shown in figure 8 of the latter.

\begin{figure}[]
\begin{center}
\includegraphics[width=9cm]{{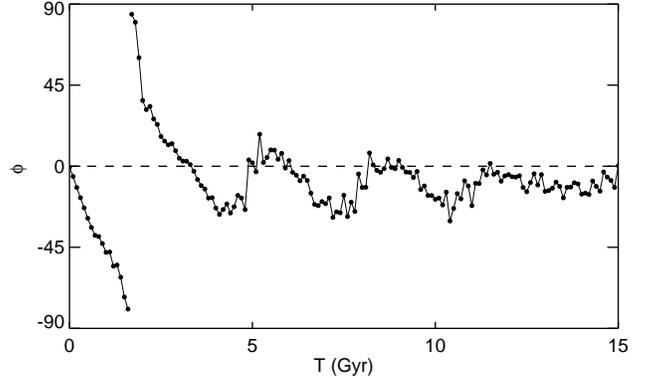}}
\caption[$\phi$ vs Time for a Triaxial DM Halo on a Circular Orbit]{\label{fig:cirdmphi}\footnotesize{Radial Alignment, $\phi$, vs time for a triaxial halo on a circular orbit. $\phi =$\degrees 0 means the major axis of the halo is pointing in the direction of the cluster center, whereas \degrees {90} means it is tangential to it. }}
\end{center}
\end{figure}

\section{The Dark Side: A Single Component Dark Matter Halo}
\label{sec:dmonly}
We begin by running a few simulations with a single component dark matter halo only, in order to directly compare the results from this idealized setup with the cosmological simulations analyzed in PBG08. The methods used to create this model are essentially the same as for the dark matter component of the full galaxy model, except that the stellar potential is not included in the Eddington inversion procedure outlined in \S \ref{sec:edd}. The shape and orientation of the dark matter halo in this section is always calculated from the reduced inertia tensor of particles (as described in \S \ref{sec:shape3d}) that are within $r \lesssim r_{\rm tid,cir} \approx 80$ kpc of the halo center, unless specified otherwise in the text.

\subsection{Circular Orbit}

As a first step we take the simplest geometric case and place the triaxial halo on a circular orbit, with the major axis initially pointing towards the cluster center, and the minor axis pointing out of the orbital plane. Figure \ref{fig:cirdmphi} shows the evolution of the radial alignment angle, $\phi$ with time. Strikingly, after one orbital period ($T \approx 5.2$ Gyr), the halo is already very close to perfect tidal locking. The halo appears to oscillate about a stable position and the amplitude of these oscillatons dampens with time, so that by the third orbit, the halo is almost perfectly aligned.

\begin{figure}[]
\begin{center}
\includegraphics[width=9cm]{{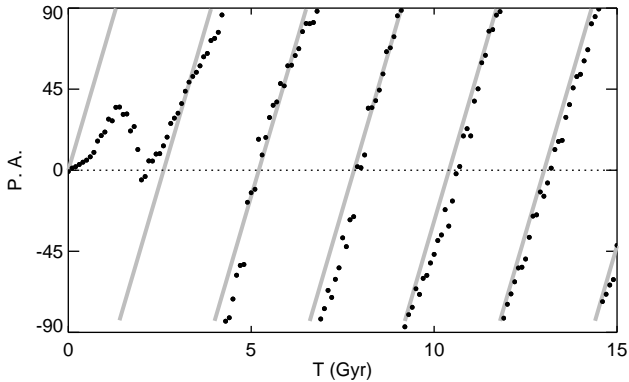}}
\caption[P.A. vs Time for a Triaxial DM Halo on a Circular Orbit]{\label{fig:cirdmpa}\footnotesize{Position angle (black circles)  and radial direction, $\theta_r$, (gray straight lines) vs time for a triaxial halo on a circular orbit. After the first orbit ($t=5.2$ Gyr), the halo undergoes figure rotation at the same rate as its orbital angular motion so that the two vectors are constantly aligned.}}
\end{center}
\end{figure}

This is shown more effectively in figure \ref{fig:cirdmpa}, where we plot both the position angle of the halo and the angle towards the cluster center, $\theta_r$. At first, the halo is torqued toward the cluster center, but by the time the halo has reached the first quarter of the orbit, at $t \approx 1.3$ Gyr, it has not rotated far enough to match the angular rotation of the orbit, and so it is torqued back to its initial  P.A. $\approx$ \degrees 0  alignment. On the second half of the orbit, this does not happen and the halo is continually torqued in the same direction. This sets up a uniform rotation throughtout the rest of the halo's evolution, where the halo shape is figure rotating at the same angular frequency of the orbital motion.

\begin{figure}
\begin{center}
{\includegraphics[width=9cm]{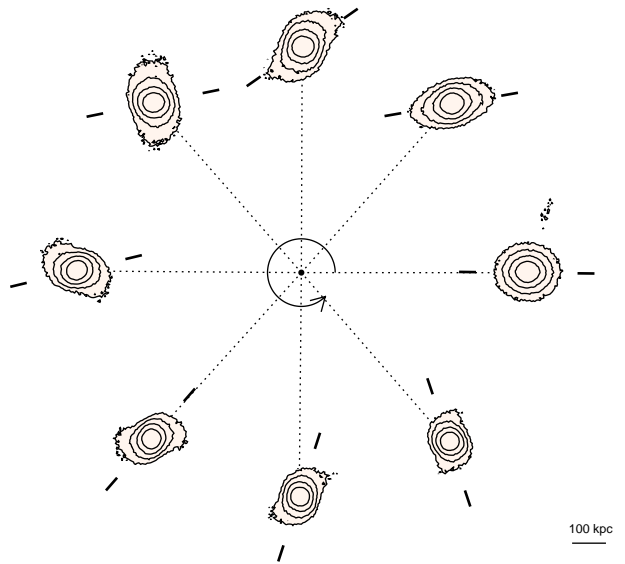}}
\caption[Density Contours of Triaxial DM Halo on Circular Orbit (1st Orbit)]{\label{fig:orb0}\footnotesize{Projected particle density contours onto orbital plane for first passage on a circular orbit. Contours are logarithmically spaced at a constant projected density. Dashed lines indicate direction to cluster center, and thick lines indicate the major axis of the halo as determined by the shape fitting procedure outlined in \S \ref{sec:shape3d}. The sequence starts with the snapshot at the far right, and proceeds counter clockwise, with each snapshot $\approx 0.65$ Gyrs apart. [Orbit is not to scale!]}}
\end{center}
\end{figure}

A visual representation of the halo's alignment is given in figure \ref{fig:orb0} for the first orbit, where the swinging for the first half orbit is evident. There is also a significant amount of isophotal twisting apparent in the contours. This is most in evidence in the outermost contour, which represents a very loosely bound component, but it is also seen closer to the center, particularly for the first half orbit. The trends of figures \ref{fig:cirdmphi} and \ref{fig:cirdmpa} do not depend sensitively on where we place the outermost boundary for the inertia tensor measurement, but we will analyze this issue more carefully in \S \ref{sec:twist}.  We note that the galaxy begins with about 20\% of the particles outside the tidal radius and loses mass as $t^{-0.2}$, with about 50\% of the mass becoming unbound after 20 Gyr, consistent with previous work \citep[e.g., ][]{Hayashi:2003p541}.

\vspace{0.5cm}
\subsection{Eccentric Orbit}

On an eccentric orbit, the magnitude of the tidal field will vary significantly. For an NFW density profile, the tidal force, $F_t$ increases monotonically with decreasing radius. At large radii, $r>>r_s$, the increase is drastic ($F_t \propto r^{-3}$). As the central regions of the halo are approached, the tidal force continues to increase, but more slowly: at $r\sim r_s, F_t \propto r^{-1}$. Further in, within $r \lesssim 0.1r_s$, $F_t$ asymptotes to a constant value.

At the same time, the halo's orbital speed is no longer constant. Halos will spend most of their time close to apocenter, and pass through pericenter very quickly. The more eccentric the orbit, the more extreme the variation in orbital speed. The angular velocity is also not constant, so that the tidal torque from the host will vary azimuthally very rapidly through pericenter, but much more slowly through the rest of the orbit. How will this affect a triaxial halo's orientation?

\begin{figure}
\begin{center}
{\includegraphics[width=9cm]{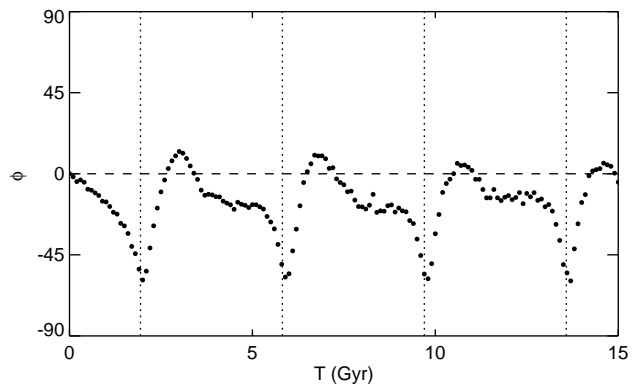}}
\caption[$\phi$ vs Time for a Triaxial DM Halo on an Eccentric Orbit]{\label{fig:dmeccphi}\footnotesize{Radial Alignment, $\phi$, vs time for a triaxial halo on an eccentric orbit. Dotted vertical lines represent pericenter passages.}}
\end{center}
\end{figure}

Figure \ref{fig:dmeccphi} shows the evolution of radial alignment for four orbital periods. The halo is mostly aligned throughout its orbit, except at pericenter (the vertical dashed lines in figure  \ref{fig:dmeccphi}) where there is strong misalignment with the radial direction. Immediately after pericenter the radial angle decreases again, and swings back toward zero, overshoots, before coming back to settle at a low offset for the rest of the orbit. This behaviour is replicated almost exactly for each orbit, and the halo shows no sign of asymptoting to a different alignment state, in contrast to the behaviour on the circular orbit, where after one orbital period, near perfect locking was achieved.

The misalignment at pericenter occurs very suddenly, and it is somewhat surprising how quickly alignment is reinstated. This trend was also observed for the average stacked population of halos in PBG08. However, our resolution was not high enough to observe the detailed dynamics for any individual halo. Naively, one would expect the effect of torquing to be largest at pericenter, since this is where the tidal field is strongest, but this overlooks the effect of the variable orbital speed, which must also be taken into account.

\begin{figure*}
\begin{center}
\includegraphics[angle=90,width=0.9\textwidth]{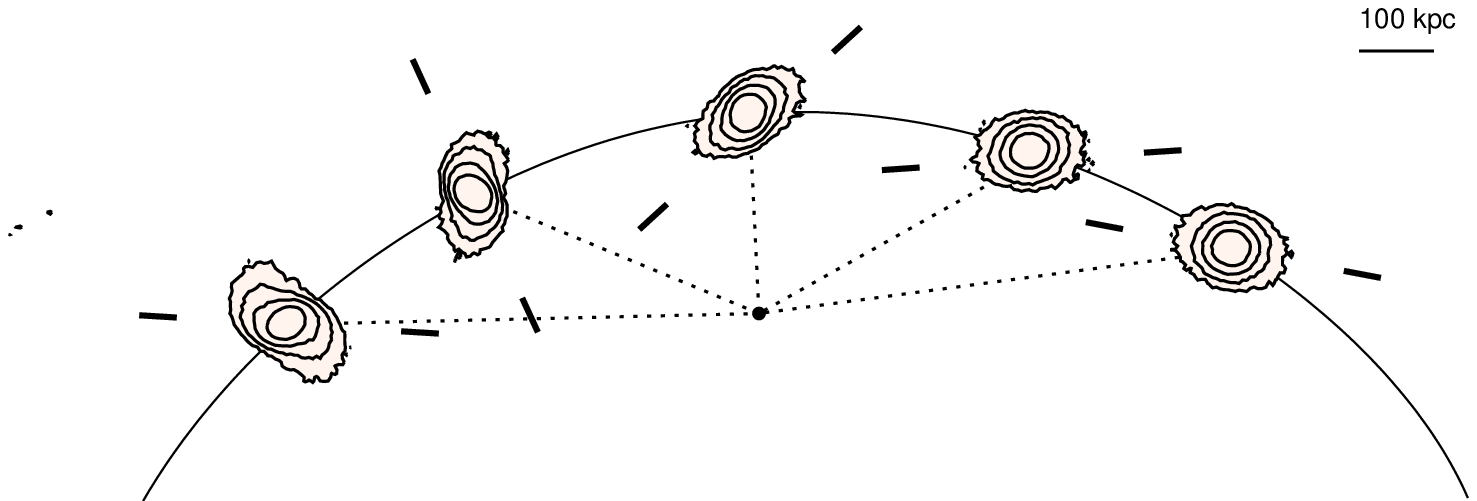}
\caption[Tidal Locking on an Eccentric Orbit]{\label{fig:orb0ecc}\footnotesize{Pericentric passage on a 4:1 eccentric orbit [Orbit not to scale]. Contours are logarithmically spaced at a constant projected density. Dashed lines indicate direction to cluster center, and thick lines indicate the major axis of the halo as determined by the shape fitting procedure outlined in section \ref{sec:shape3d}. }}
\end{center}
\end{figure*}

In PBG08 we suggested that the sudden misalignment could be due to the very high orbital speeds during each pericentric passage. The halos are moving too fast to be efficiently torqued, and therefore the radial angle, $\phi$, increases. Soon after pericenter the alignment is recovered, and the halo is torqued back onto the radial direction, in effect swinging around back onto itself. These high resolution experiments have confirmed this, and allow us to actually \emph{see} this effect in action. Figure \ref{fig:orb0ecc} shows the projected particle density contours onto the orbital plane for the second pericentric passage on a 4:1 orbit.  The halo travels in a counter-clockwise direction from right to left and we show here 5 snapshots separated by 0.3 Gyrs.

\subsection{Internal Structure and Shape Twisting}
\label{sec:twist}

Figures \ref{fig:orb0} and \ref{fig:orb0ecc} show some amount of twisting in their isodensity contours. Certainly at larger radii, beyond the tidal radius of the satellite, one expects the cluster's tidal field to distort the shapes of the more loosely bound particles, eventually leading to the formation of tails and streamers. What happens closer to the center of the satellite? Is the internal triaxial structure of the halo perturbed by the tidal torquing, or does it maintain its shape through its orbit?

\begin{figure}
\begin{center}
{\includegraphics[width=9cm]{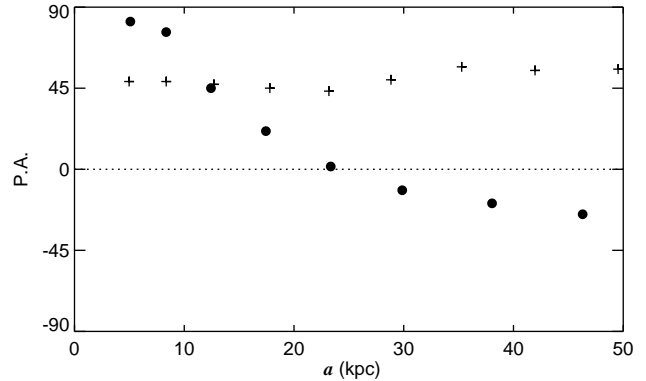}}
\caption[Isophotal Twisting Profile]{\label{fig:twist}\footnotesize{Position angle vs $a$, the isophotal semi-major axis. Pluses are for the halo at apocenter ($T=3.9$ Gyr), and dots are for the halo immediately after pericenter ($T=5.9$ Gyr). While at apocenter there is almost no twisting in the projected mass profile  $\Delta \phi <$ \degrees{15}, at pericenter the twisting is extreme, $\Delta$ P.A.$ >$ \degrees{100}.  }}
\end{center}
\end{figure}

We explore this in figure \ref{fig:twist}, which shows the position angle of the satellite as a function of satellite radius. More precisely, the orientation of the satellite is measured at different radii by fitting ellipses (as described in \S 2.9 ) to logarithmically spaced projected mass contours, with the corresponding semi-major axes plotted in the figure. The position angle profile is shown for the satellite at the two orbital extremes, apocenter (pluses) and pericenter (dots). At apocenter there is almost no twisting observed ($\Delta$ P.A. $ <$ \degrees{15}), while, at pericenter, the shape twisting is extreme ($\Delta$ P.A. $ >$ \degrees{100}), even though the position angle varies smoothly with radius, with no obvious break within this radial range.

Remarkably, this shape twisting is seen even at small radii, well within the scale radius of the halo ($r_s\approx 23$ kpc). It seems likely that the compact stellar component of the galaxy will react to the external torque in much the same way as the inner particles of the dark matter halo. The stars will therefore be misaligned with the outer halo, particularly at pericenter, and this must play a part in the discrepancy between the alignment seen in observations and that of simulated cosmological halos. We will explore this further in section 4.
 
\section{The Bright Side: Adding a Luminous Stellar Component}
\label{sec:st}

The large number of particles in these simulations has allowed us to delve deeper into the inner regions of the dark matter halo, and we have confirmed that, for a halo which is initially triaxial, the whole halo undergoes approximately rigid body rotation through most of its orbit, with the exception of pericenter passages which induce significant shape twisting. However, the PK05 observations are made on much smaller scales, within 2 or 3 luminous effective radii. At this numerical resolution, the number of dark matter particles within this radius may not be enough to adequately sample the dynamics of this region. Moreover, dynamical models of elliptical galaxies suggest that the phase space distribution of stars differs from that of the dark matter halo. A more complex model which includes a centrally concentrated stellar distribution is required, and this will also allow us to make more direct comparisons between our model galaxies and their observed shapes.

Once stellar particles are included in our model, we can start simulating the observing process itself. In that spirit, instead of measuring average 3D shapes for our galaxy we will use isophotal contours to measure galaxy shapes at the same physical radius as the SDSS galaxies observed in PK05 ( $I_{25} = 25$ mag/arcsec$^{2}$, corresponding to $r_{25} \approx 10$ kpc, see \S \ref{sec:shape3d} for a description of the method). In addition, we choose to measure an equivalent isodensity contour for the dark matter, since this makes the comparison between the two components easier, and is also physically motivated by the possibility of observing projected mass maps through indirect methods such as lensing techniques. The isodensity contour for the dark matter is measured at $I_{\rm DM} =  10.7\times10^6  M_{\sun}$ kpc$^{-2}$, which results in an average contour radius $<r_{\rm DM}> \approx 40$ kpc. The orientations calculated using this method are close to those obtained using the 3D inertia tensor, although not exactly the same, since they measure shapes at a specific radial distance as opposed to a mean shape for the entire halo.

\begin{figure}
\begin{center}
{\includegraphics[width=8.5cm]{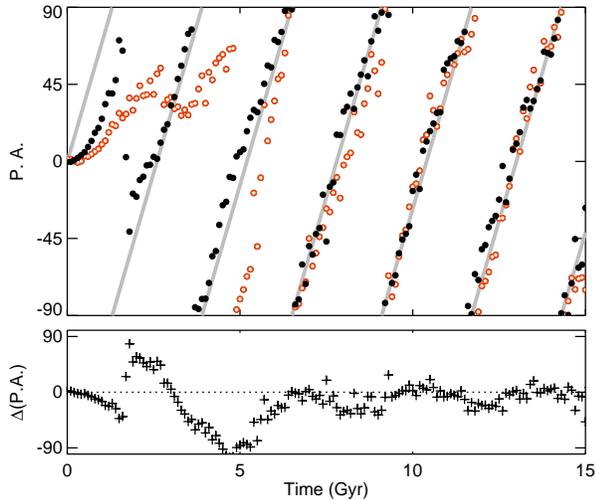}}
\caption[Position Angle for Triaxial Galaxy in Circular Orbit as a function of Time]{\label{fig:orbcir_stdmr}\footnotesize{Top: Position angle for triaxial galaxy on a circular orbit as a function of orbital time. The dark matter component (dark filled circles) responds more quickly to the tidal torque than the stellar component (open light circles) although after 1.5 orbits their behavior is indistinguishable. Bottom: The misalignment between the dark matter and stellar orientation, $\Delta$P.A. as a function of orbital time.}}
\end{center}
\end{figure}

\subsection{Circular Orbit} 

\label{sec:stcir}

We saw in the previous section that a triaxial dark matter halo on a circular orbit around a cluster potential becomes tidally locked after just one orbit. What happens to the stellar particles that reside at the very centers of these dark matter wells? The stellar particles have a different phase-space distribution and may therefore react differently to the same external tidal field. On the other hand the stars and dark matter are not completely decoupled: dynamical friction acts between the two components and will work to bring the two components into alignment \citep{Nelson:1995p897}. 

Using exactly the same orbital setup, we repeat the experiments of the previous section, but now with the full galaxy model. Figure \ref{fig:orbcir_stdmr} shows the measured position angles of both the stellar and dark matter components of our model galaxy as a function of time in orbit. The first thing to note is that after 6 Gyrs the behavior of the two components is indistinguishable: they both show figure rotation in the orbital plane that matches the orbital period exactly. There is a significant difference between the two in the first orbital passage however, with the dark matter halo responding much more quickly to the external applied torque. The lag between stellar alignment and dark matter alignment can be almost a full orbital period. Since massive clusters have recent formation times, many of their members will have fallen in less than 5 Gyrs ago, and this will be an important dilution to the radial alignment signal. This lag in alignment will be explored in more detail in the coming sections.

It is important to note that, if measured at the same radius, e.g. $r_{25}$,  the orientations of the dark matter particles and stellar particles are very similar, and always within $ \lesssim $ \degrees {15}, i.e. the induced misalignment between the dark matter and stellar alignment is mostly driven by the fact that the two are measured at different radii (each radius being representative of the spatial extent of the component). A misalignment between the stellar component at $r_{25}$ and the dark matter component at $r_{\rm DM}$ is also a misalignment between the inner and outer dark matter halo. The orientation of the dark matter halo appears to change smoothly between the inner and outer regions, as is evident in the shape twisting shown in figure \ref{fig:twist}.

\begin{figure}
\begin{center}
{\includegraphics[width=9cm]{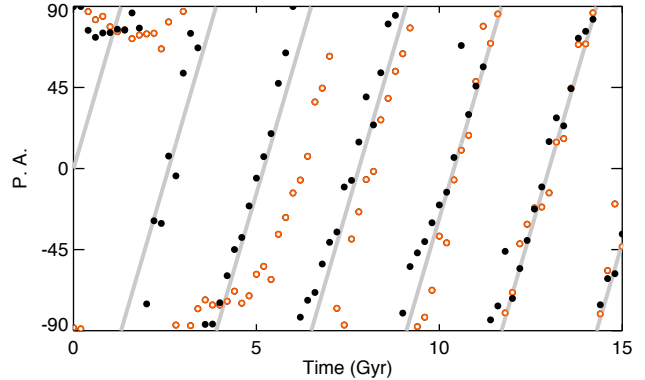}}
\caption[Position Angle for Triaxial Galaxy in Circular Orbit as a function of Time with Initial Tangential Orientation]{\label{fig:orbcir_stdmt}\footnotesize{Position angle for triaxial galaxy on a circular orbit as a function of orbital time, with initial tangential alignment. Symbols as in the previous figure. If the galaxy is initially tangentially aligned to the cluster center, tidal locking takes a longer time to set in, but after 1.5 orbits, or approximately 7.5 Gyrs, both components are tidally locked in radial alignment. The dark matter component spins up much faster than the stellar particles: While the dark matter rotates a full \degrees {360} in the first orbit,  the stars are slower to react to the external torque, and have only rotated by $\approx$ \degrees {180}. }}
\end{center}
\end{figure}

\subsubsection{Dependence on Initial Galaxy Alignment}

All our simulations so far have been initialized with the galaxy's major axis along the radial direction, the intermediate axis along the orbital direction, and the minor axis out of the orbital plane. This is not an unreasonable assumption, since, as we found in PBG08, galaxies are already preferentially radially aligned when they fall into clusters along filaments. Nevertheless, in a realistic scenario the initial alignment is not perfect and so it is important to understand how sensitive our results are to this initial orientation.


We take here the extreme case and start one of our galaxies with the major and intermediate axes swapped, so that the galaxy's position angle is now tangential to the radial direction. The minor axis is still normal to the orbital plane. Both stellar and dark matter components initially rotate clockwise, as a response to the tidal torque  (figure \ref{fig:orbcir_stdmt}). However, for both components, this motion comes to a stop and is soon reversed. The dark matter aligns more quickly than the stars, with the dark matter figure rotation accelerating rapidly to match the orbital period within 2 Gyrs. The stellar component takes longer to react to the applied torque - its position angle initially rotates more slowly in the first few Gyrs, although it eventually accelerates to match the orbital motion within 7 Gyrs or 1.5 orbits.

\begin{figure}
\begin{center}
{\includegraphics[width=9cm]{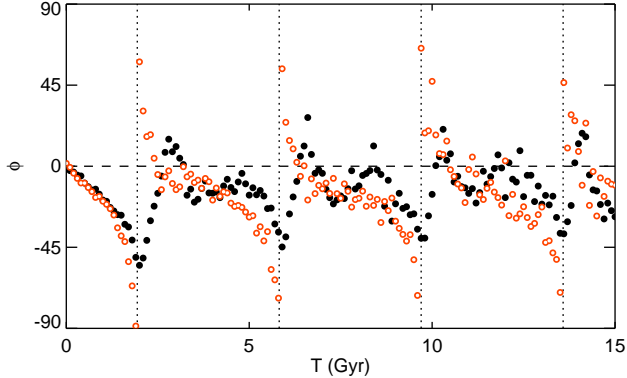}}
\caption[Radial Alignment on a 4:1 Eccentric Orbit]{\label{fig:triecc4_radstdm}\footnotesize{Radial angle for both stellar and dark matter component, symbols as before. The two components exhibit similar behavior throughout most of the orbit, except at pericentric passages, where they respond very differently to the sudden changing torque.}}
\end{center}
\end{figure}

\subsection{Eccentric Orbits}

\label{sec:stecc}

It is interesting to consider what happens to luminous galaxies on eccentric orbits, which are more typical of the cluster environment. As we saw with the single component dark matter halo, on an eccentric orbit, permanent perfect radial alignment is unachievable, since the torque direction and strength vary too quickly along the orbit. The dark matter halo tends to respond quite quickly to the varying torque, except right at pericenter, where there is significant misalignment, leading to a strong swing back immediately after.

What happens to the stars in this galaxy? Figure  \ref{fig:triecc4_radstdm} shows us that, again, their behavior is very similar in all parts of the orbit except at pericenter, where, while the dark matter gets torqued back towards the cluster, the stars respond very differently during and immediately after pericenter.

\begin{figure}
\begin{center}
{\includegraphics[width=9cm]{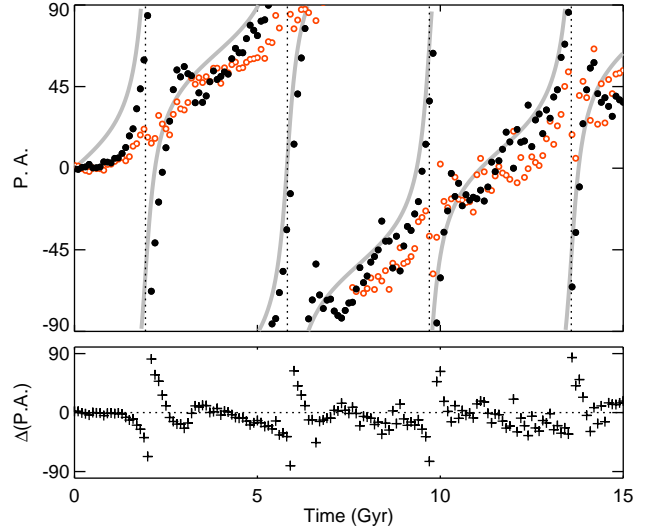}}
\caption[Stellar and Dark Matter Position Angles for an Eccentric Orbit]{\label{fig:ecc4trist90s1}\footnotesize{Top: Stellar and dark matter position angles vs time for an eccentric orbit. Gray lines indicate the direction to the cluster center. While the dark matter particles appear to follow the orbital motion closely, varying their rotation speed to match the orbital velocity changes, the stellar particles appear to be locked in uniform figure rotation. Bottom: The misalignment between the dark matter and stellar orientation, $\Delta$P.A. as a function of orbital time. In contrast with the circular orbit case, the misalignment here does not asymptote to zero, but is instead correlated with the orbital phase:  $|\Delta$P.A$|$ peaks immediately after each pericenter passage, and remains small for the rest of the orbit, indicating internal alignment of the galaxy.}}
\end{center}
\end{figure}

This can also be clearly seen when we plot the position angle of the isophotal contour vs time in figure \ref{fig:ecc4trist90s1}. Whereas the dark matter isophote follows the orbital motion rather closely, the stellar component essentially figure rotates at a steady rate, with the position angle changing linearly with time. This rate is maintained even through pericenter (especially in the first passage), when the tidal field is changing very rapidly.

\begin{figure*}
\begin{center}
{\includegraphics[width=\textwidth]{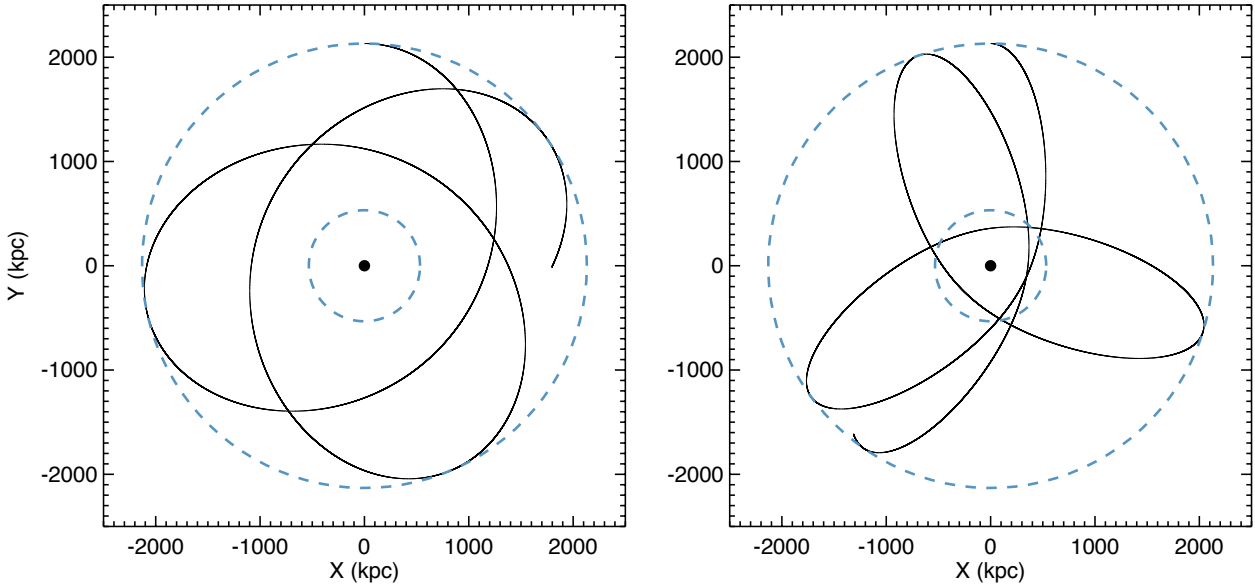}}
\caption[2:1 and 6:1 Orbital Paths]{\label{fig:ecc62}\footnotesize{Galaxy trajectories in 2:1 (Left) and 6:1 (Right) eccentric orbits.}}
\end{center}
\end{figure*}

\begin{figure*}
\begin{center}
{\includegraphics{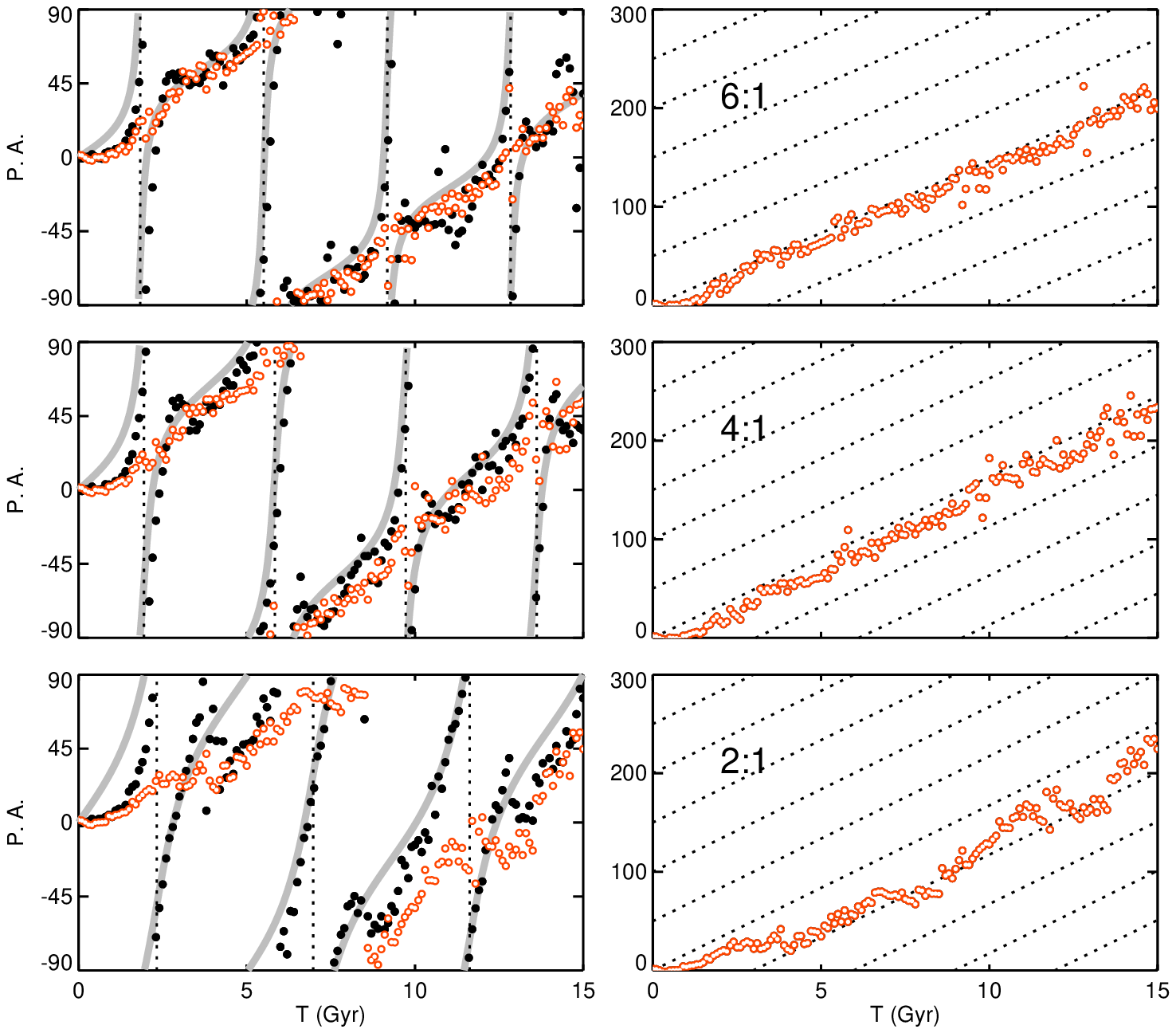}}
\caption[Galaxy Position Angles for Orbits with Varying Eccentricities]{\label{fig:triecc_all}\footnotesize{Stellar and dark matter position angles vs time for 6:1 (top), 4:1(middle) and 2:1 (bottom) orbits. Left panels show the radial direction in gray lines and are folded at \degrees {90} and \degrees {-90}. Right Panels show the running stellar position angle, with the dashed lines representing the tumbling frequency, $\Omega_{ gal}$ calculated with equation \ref{eq:freqs}}}
\end{center}
\end{figure*}

\begin{figure*}
\begin{center}
{\includegraphics{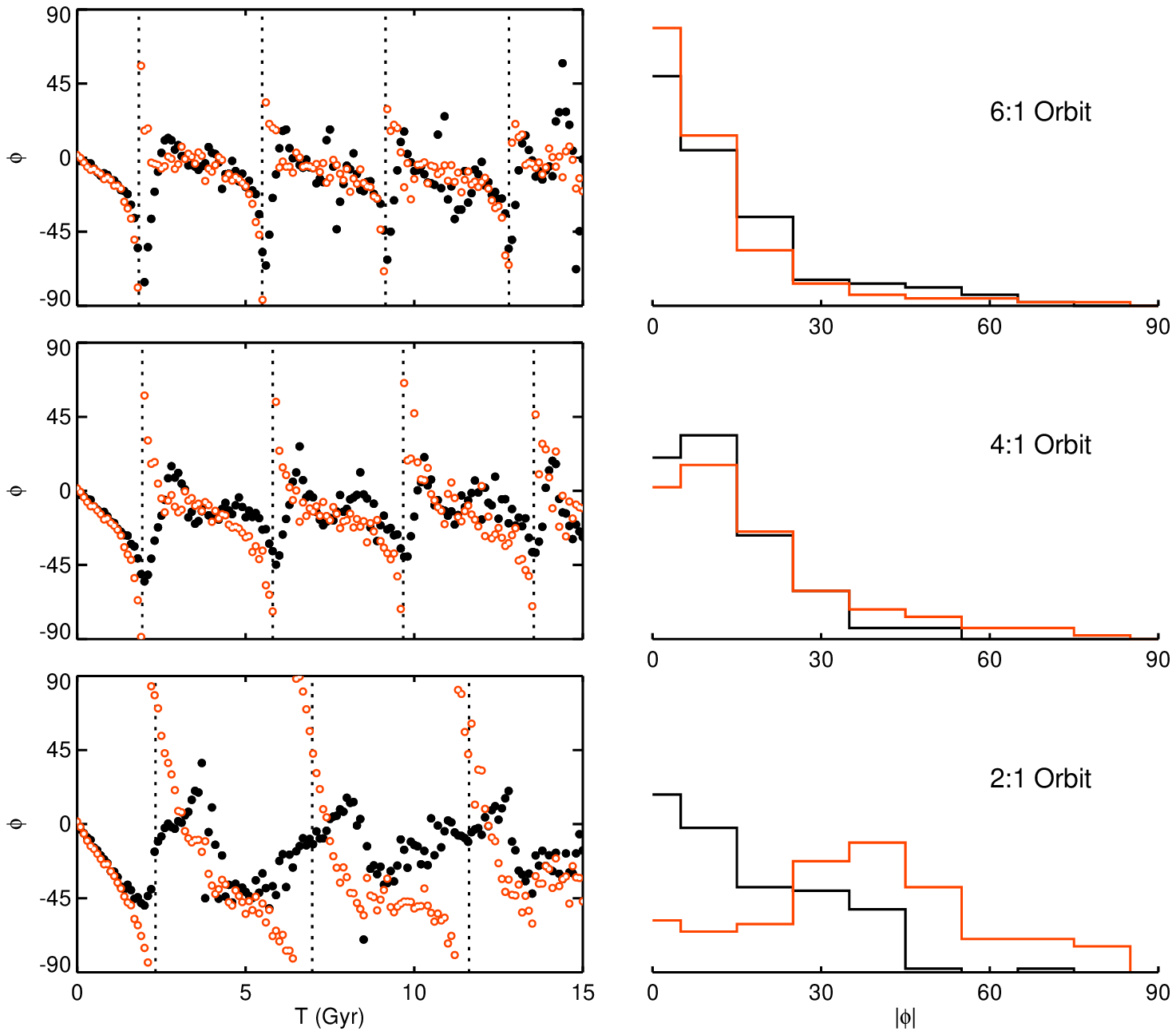}}
\caption[Galaxy Radial Angles and Histograms for Orbits with Varying Eccentricities]{\label{fig:triecc_radall}\footnotesize{Left Panels: Stellar and dark matter radial angles vs time for 6:1 (top), 4:1(middle) and 2:1 (bottom) orbits. Right Panels: Histograms of relative frequencies of $\|\phi\|$ for the dark matter and stellar components}}
\end{center}
\end{figure*}

Is this a consequence of the specific geometry of this 4:1 orbit, or a generic feature of satellites orbiting in NFW potentials? Figures \ref{fig:triecc_all} and \ref{fig:triecc_radall} show the outcome of this analysis for three different orbits, each with the same apocenter, but apocenter to pericenter ratios of  6:1, 4:1 and 2:1 (see the orbits in figure \ref{fig:ecc62}). It is evident from these plots that the 4:1 case was not unique. Certainly for orbits which are even more eccentric, the figure rotation induced in the stellar particles is even more uniform, and radial alignment more frequent.

This figure rotation frequency, $\Omega_{\rm gal}$, appears to be the lowest rotational frequency that minimizes the energy of the system (by maximizing the mean radial alignment), and it can be estimated from the difference between the azimuthal ($\Omega_{\psi} $) and radial ($\Omega_r$) frequencies of the cluster orbit: 
\begin{equation}
\label{eq:freqs}
\Omega_{\rm gal} = \Omega_{\psi} -{\Omega_r}/{2}~.
\end{equation}

This behavior is reminiscent of a spin-orbit resonant interaction, in that the frequencies of the orbits and spins satisfy a commensurability condition. The frequencies derived from this equation are used to produce tumbling tracks which are shown on the right panel of figure \ref{fig:triecc_all} as dashed lines. For the top two panels, the tumbling slopes are a remarkable fit. The rotation of the galaxy's position angle is smooth and uniform, as if the galaxy is not aware of the abrupt change in the tidal field arising from the high eccentricity of the orbit. 

In this regime, the tumbling frequency, $\Omega_{\rm gal}$, is close to the angular velocity at apocenter, $\frac{d\psi}{dt}_{{\rm apo}}$, but it is much, much smaller than the angular velocity at pericenter,$\frac{d\psi}{dt}_{{\rm peri}}$. The stars cannot react to the rapid variation in the tidal field at pericenter, responding instead to the mean torque throughout the orbit.

As the eccentricity decreases, the mean azimuthal frequency remains approximately constant, but the variation in  $\frac{d\psi}{dt}$ along the orbit becomes smaller: $\frac{d\psi}{dt}_{{\rm peri}} / \frac{d\psi}{dt}_{{\rm apo}} \approx 4 $, an order of magnitude smaller than in the 6:1 orbit. $\Omega_{\rm gal}$ is now much closer to  $\frac{d\psi}{dt}_{{\rm peri}}$ and we see that indeed the pericenter passages induce significant oscillations in the position angle of the stellar component (bottom panel of figure \ref{fig:triecc_all}). This leads to an overall offset from radial alignment, which is immediately apparent in the histograms of figure \ref{fig:triecc_radall}.  

It seems, then, that orbits with intermediate eccentricities ($r_{\rm apo}/r_{\rm peri} \sim 2$) can lead to significant stellar misalignment throughout the orbit. The dark matter halo is also disrupted by pericenter passages and less radially aligned on average throughout its orbit than in the more eccentric cases, but much less so than the stars.   What is most striking in the 2:1 orbit is the overall misalignment between the stellar and dark matter particles. Simulations with apocenter to pericenter ratios of 3:1 and 3:2 bridge the results shown in figure \ref{fig:triecc_radall}: As the eccentricity of the orbit decreases, $<|\phi|>$ increases to higher values (until the orbit becomes almost circular).

\subsection{A Comparison with Observations}

These results indicate that when galaxies on orbits with varying eccentricities and different accretion times are combined in an observational sample of cluster galaxies, such as that analysed in PK05, the average radial alignment measured for the stellar component will be much lower than that inferred for the dark matter from cosmological simulations. This is a direct consequence of the shape twisting observed in figure \ref{fig:triecc_radall}: while on more eccentric orbits, the galaxies are internally aligned except at pericenter, for more circular orbits, this misalignment is present throughout most of the orbit. In the extreme case of a circular orbit (figure \ref{fig:orbcir_stdmr}), the two components eventually become aligned, but only after a couple of orbital passages - recently accreted galaxies will be misaligned. 

An evolving, non-spherical cluster potential will break the symmetry of these satellite orbits, which will affect the results in the previous section to some degree, but the general conclusions should remain unchanged. A more complete study that takes into account projection effects, evolving satellite orbits and the intrinsic noise in the observational measurements would be required for a quantitative comparison to observational results.

\section{The Impact of Tidal Torquing on Galaxy Shapes and Dynamics}
\label{sec:shapevol}

\subsection{Induced vs Intrinsic Triaxiality} 

A natural question to ask is whether the alignment toward the cluster center is a result of tidal stretching, where the alignment is induced by the tidal field, and the shape of the halo is elongated along the potential gradient, or more correctly described as a tidal torquing effect, where the halo, which is intrisically triaxial, can be treated approximately as a solid body which can be torqued about its minor axis while maintaining its original shape. We will see that in fact, neither simplifying description is adequate, and a better physical description lies somewhere between these two extremes.

\begin{figure}
\begin{center}
{\includegraphics[width=9cm]{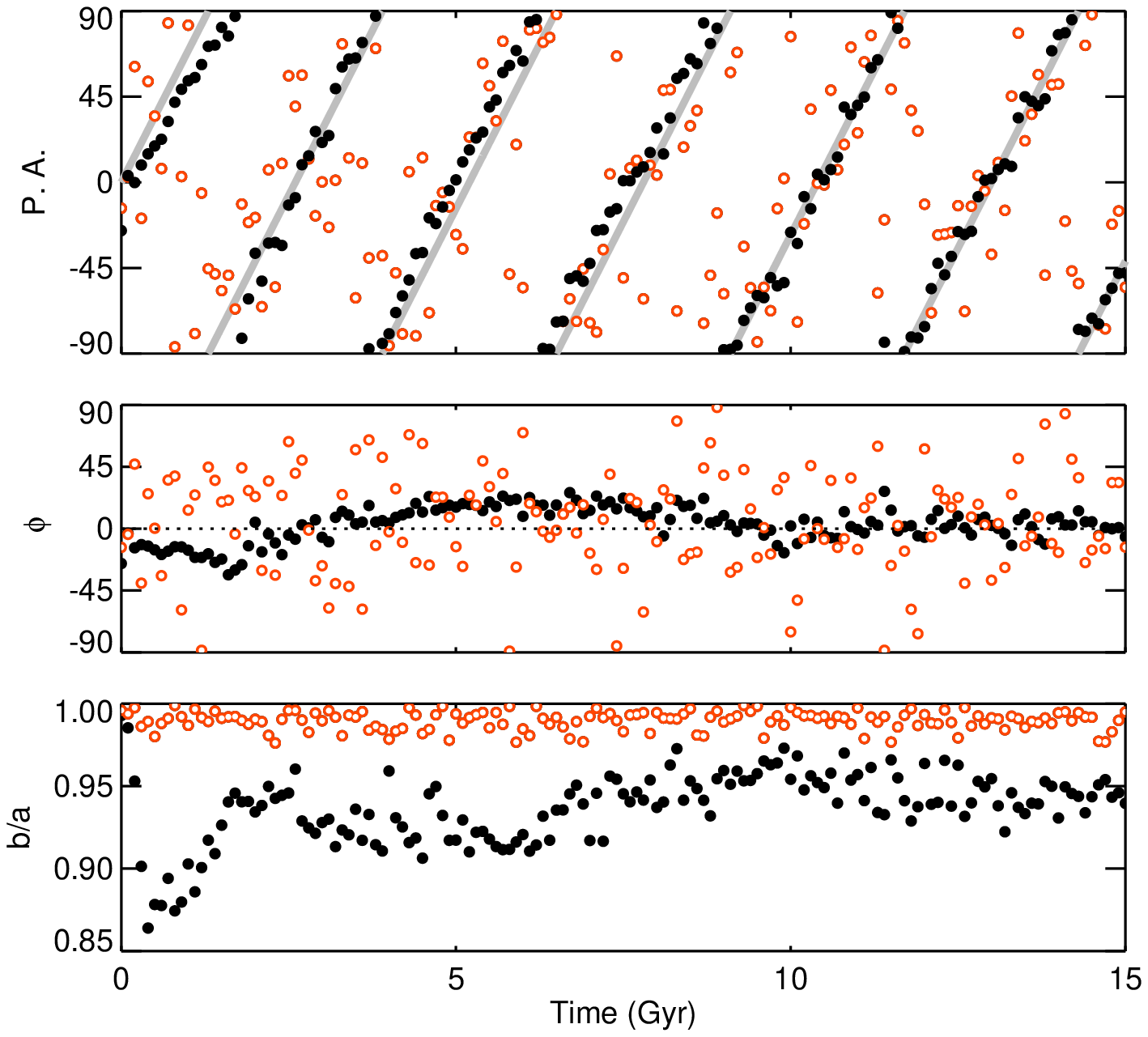}}
\caption[Spherical Galaxy on a Circular Orbit]{\label{fig:sphcir}\footnotesize{A spherical galaxy on a circular orbit. From top to bottom: The position angle, radial angle, and projected axis ratios of the dark matter (filled dark circles) and stellar components (open light circles) as a function of time. Both components are initially perfectly spherical, but while the stellar component retains its shape, the more extended dark matter distribution is sheared by the tidal field and becomes significantly elongated. This elongation is at all times in the direction of the cluster center, so that the DM component is essentially in perfect radial alignment from $t=0$. The stellar component orientation appears random, as is expected for a spherical distribution.}}
\end{center}
\end{figure}

As a first test, we analyze the shape evolution of an initially spherical galaxy which is placed on a circular orbit around the cluster. In the bottom panel of figure \ref{fig:sphcir}, we see that while the projected axis ratio of the stellar component remains very close to spherical ($b/a \approx 1$), the dark matter halo's shape is significantly altered by the external field. As a result of this, the dark matter halo shows instant radial alignment (since the elongation caused by the tides is always along the potential gradient), and it maintains this induced alignment throughout its orbit. In contrast, the shape of the stellar particle distribution, being much more tighly bound, is mostly unaffected by the tidal field, and no radial alignment is induced.

A simplifying, but useful, description is to analyze the system in terms of its characteristic frequencies. The outer particles in the galaxy orbit the galaxy at approximately the same frequency as the galaxy orbits the cluster, since, by definition, the mean density of the satellite within the tidal (or Roche) radius is close to the mean density of the cluster within the galaxy's orbital radius ($\bar{\rho}_{gal}=3\bar{\rho}_{cluster}$). For these particles, the timescale for the azimuthal variation in the tidal field is comparable to their own orbital frequencies, which means that their orbital structure will change significantly as a reaction to the external torque.  

At the other extreme, the stellar particles at the very center of the galaxy orbit at much higher frequencies - to these particles, the variation in the external tidal field is essentially an adiabatic process. In a slowly varying potential, the action integral $J_i =  \int p_i {\rm d}q_i$, where $p_i$ are the conjugate momenta and $q_i$ are the generalized coordinates, is an adiabatic invariant, a conserved property of the particle orbit (Binney \& Tremaine 1987). In this limit, the orbital structures that preserve the galaxy's triaxial shape are unaffected by the slowly varying external torque, and the galaxy will react to the torque as a rigid body, figure rotating around its minor axis in response to the outside torque. Of course, most particles in our galaxy are in an intermediate regime, where some of the energy from the tidal torque will go into heating the particle orbits, and some into a coherent rotation of the triaxial shape. Evidence of this is seen in the evolution of the galaxy's shape, which we discuss in the next section. It is important to note that a full particle orbital analysis would give us a better understanding of how individual orbit families react to the torquing, and should be undertaken in future work.

\subsection{Shape Evolution in Substructure} 

\begin{figure}
\begin{center}
{\includegraphics[width=9cm]{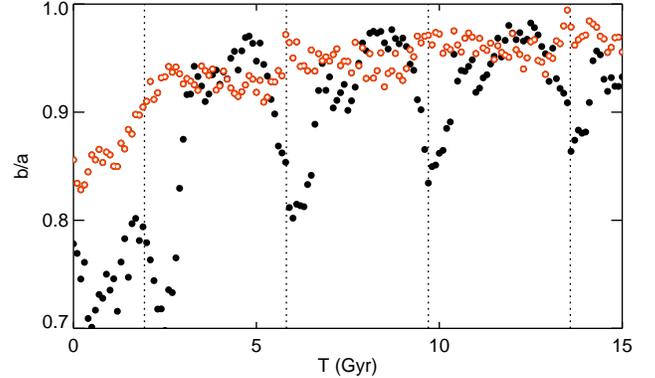}}
\caption[Evolution of Galaxy Axis Ratios with Time for the 4:1 Eccentric Orbit]{\label{fig:ecc4mtri_evals}\footnotesize{Evolution of axis ratios with time for the 4:1 orbit for dark matter (dark circles) and stellar particles (open red circles). At each pericenter passage (vertical dashed lines), the dark matter becomes quite elongated, whereas the stellar component becomes more spherical.}}
\end{center}
\end{figure}

Figure \ref{fig:ecc4mtri_evals} shows the temporal evolution of the triaxial galaxy shape on the 4:1 orbit around the cluster, and its correlation with orbital phase. The dark matter becomes quite elongated at pericenter passages, as is expected for particles that are loosely bound: at pericenter the tidal radius for the galaxy shrinks momentarily down to $r_t \sim 50$ kpc. At the same time, pericenter passages make the stellar component rounder. This is probably mostly due to tidal shocking. The quick change in the potential around pericenter causes orbital mixing, heating the stellar orbits, which will puff the galaxy up and make it more spherical. But what drives the increase in the stellar component axis ratio for the first 2 Gyrs? The shape of the galaxy is stable in isolation, and the cluster potential is turned on adiabatically, so the loss of ellipticity must be somehow associated with its orbital motion.

\begin{figure}
\begin{center}
{\includegraphics[width=9cm]{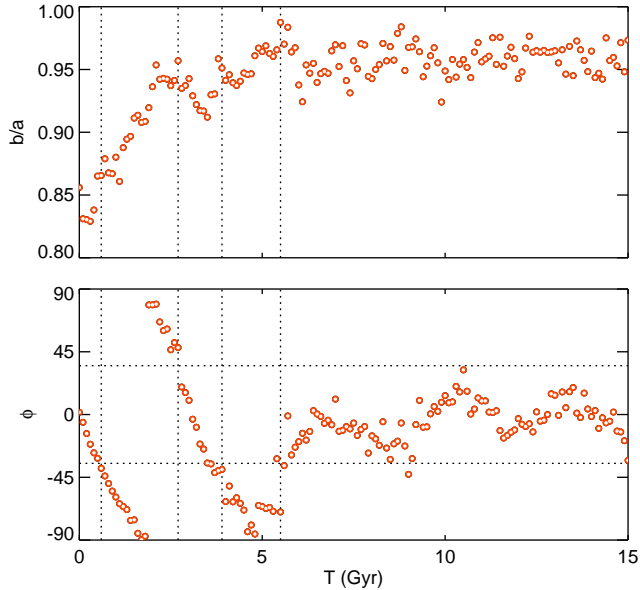}}
\caption[Shape Evolution vs Radial Alignment for a Circular Orbit]{\label{fig:ecc4mtri_evalscir}\footnotesize{ Relation between shape evolution (top) and radial alignment (bottom) for the stellar component of a galaxy on a circular orbit. Vertical dashed lines mark the transition between regimes of large ($\|\phi\| >$\degrees {35}) and small radial misalingment  ($\|\phi\| <$\degrees {35}).  The figure suggests that periods of large radial misalignment ($\|\phi\| >$\degrees {35}) lead to circularization of the galaxy shape. }}
\end{center}
\end{figure}

This becomes clear when we look at the temporal evolution of axis ratios for a circular orbit. In this case, there is no tidal shocking, the magnitude of the tidal torque is constant, and its direction varies uniformly with time. Nevertheless, figure \ref{fig:ecc4mtri_evalscir} shows similar evolution towards larger axis ratios in the beginning of the simulation, and another short period of circularization about 5 Gyrs into the orbit.

By comparing this behavior with the bottom panel of figure \ref{fig:ecc4mtri_evalscir} we can see that in fact the episodes of circularization occur only when the galaxy is significantly radially misaligned ($\|\phi\| >$\degrees {35}). At these points the torque is strongest, and while it does induce figure rotation, bringing the entire galaxy back into alignment within a couple of Gyrs, it also will isotropize the orbits slowly and make the galaxy rounder, since the adiabatic approximation is not perfect.

This trend has in fact been detected by \citet{Kuehn:2005p1032} in a study of galaxy shapes in clusters using SDSS. They find that the axis ratios of elliptical galaxies tend to become rounder in denser environments, even after correlations with intrinsic luminosity are accounted for. 

\subsection{The importance of initial triaxiality}

\begin{figure}
\begin{center}
{\includegraphics[width=9cm]{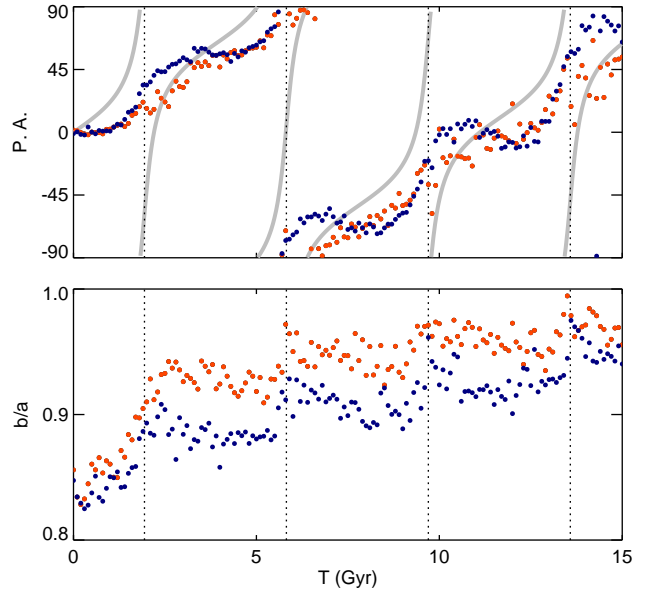}}
\caption[Comparison of the torquing of two halos with different degrees of triaxiality.]{\label{fig:ecc4mtri_two}\footnotesize{Tidal torquing of two halos with different degrees of triaxiality. The top panel shows the position angle of the two galaxies overlayed on $\theta_r$, the angle in the direction towards the cluster center (in gray). The bottom panel shows the projected axis ratios of the two galaxies on the orbital plane. Dashed lines signal pericenter passages. Red dots are our fiducial galaxy, and blue dots are the more triaxial model discussed in the text.}}
\end{center}
\end{figure}

How does the alignment depend on the initial triaxiality of the galaxy? A galaxy which is more triaxial should be more responsive to an external torque. We test this by creating a model galaxy that is more triaxial (by increasing $T_C$ in \S \ref{sec:triax} from 2.5 to 4 Gyrs), and repeating the simulations of the previous section.  In the case of a 4:1 eccentric orbit (figure \ref{fig:ecc4mtri_two}), we see that, in fact, at each pericenter passage the position angle of the more triaxial galaxy reacts to the tidal torque more strongly. This leads to some oscillation away from the uniform rate seen in the less triaxial case, although the mean rotation rate remains the same. These oscillations grow in amplitude with each orbital passage.

Also notable is the fact that the process of sphericalization seems to be less effective overall for the more triaxial galaxy. The relative increase in axis ratios appears to be smaller (see bottom panel of figure \ref{fig:ecc4mtri_two}). This fits in with the previous picture: both galaxies experience the same tidal field at pericenter, but, while the fiducial model absorbs most of that energy into random orbital motions, increasing its sphericity, the more triaxial system reacts to the torque by figure rotating towards the cluster center, gaining coherent angular momentum.

\subsection{Isolated Evolution of a Tumbling Galaxy}
\label{sec:iso}
It seems clear by now that the tumbling motion induced in our triaxial galaxies by the external potential is not just an instantaneous tidal deformation, but rather a true torquing of the galaxy's intrinsic shape. In the case of a galaxy on a circular orbit (see, e.g. figure \ref{fig:orbcir_stdmr}) after a few Gyrs the entire galaxy becomes tidally locked, so that it tumbles over its long axis with a uniform rotation period. Once the galaxy becomes locked, is the cluster potential required to maintain the tumbling, or is the motion sustainable without the forcing of the external tidal field?
 
In order to test this, we extract one of our models (the one which is more triaxial initially) which has been orbiting the cluster on a circular orbit for 10 Gyrs, and evolve it in isolation. We begin by determining which particles remain bound to the satellite, since though the satellite's initial profile was truncated,  the torquing from the tidal field leads to continuous mass loss throughout the orbit  \citep[see, e.g., ][]{Hayashi:2003p541,Kampakoglou:2007p775} .

\begin{figure}
\begin{center}
{\includegraphics[width=9cm]{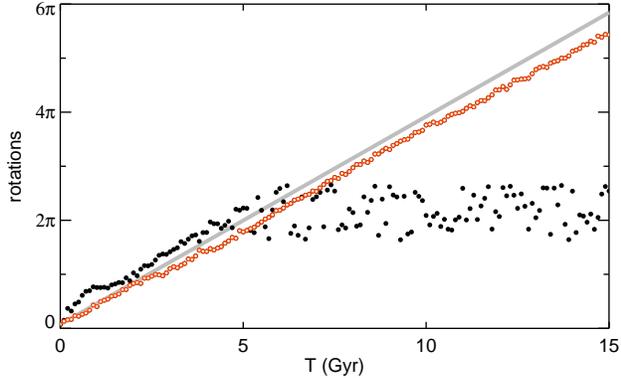}}
\caption[Triaxial Galaxy, Torqued in Circular Orbit for 10 Gyrs, Now Evolving in Isolation]{\label{fig:ecc4trist90s}\footnotesize{ Triaxial galaxy, torqued in circular orbit for 10 Gyrs, now evolving in isolation.  Both dark matter (dark circles) and stellar particles (open red circles) tumble at a close to uniform frequency in the first 5 Gyrs.}}
\end{center}
\end{figure}

The collection of bound particles is then extracted and evolved in isolation for another 15 Gyrs. Figure \ref{fig:ecc4trist90s} immediately tells us that the stellar component maintains its tumbling motion: the position angle of the galaxy increases smoothly, at a frequency which is very close to the locking frequency for a period greater than a Hubble time. The tumbling is encoded in the stellar particle orbits, and is now an intrinsic property of the satellite. There is a small sign of deceleration, since the slope of this line becomes progressively shallower, but this is to be expected, since any orbital mixing will naturally lead to a loss of coherent motion of the figure.

For the dark matter component, the case is not so clear. Initially the dark matter particles also undergo coherent rotation, but after about 5 Gyrs, or one full rotation, the shape of the distribution becomes too spherical too determine shapes reliably, and the signature of coherent figure rotation disappears. This sphericalization is seen in both dark matter and stellar particles and must be at least partly driven by the fact that the cluster potential "turn-off" was done abruptly, and such a sudden change in potential leads to shocking, isotropizing the orbits. It is nevertheless remarkable that the stellar component can maintain such a coherent rotation over such a long period of time, with only a very small sign of deceleration. 

Figure rotations of dark matter halos were measured by \citet{Bailin:2004p27} from a large cosmological simulation. They measure pattern speeds for 288 halos and find that they follow a lognormal distribution centered at $\Omega_p=0.148$ h km s$^{-1}$ kpc$^{-1}$ with a width of 0.83, and ascribe the figure rotation of their halos to tidal torques from surrounding large scale structure. Their analysis excluded all subhalos, and also halos with significant substructure or signs of recent merging, and therefore applies to somewhat isolated halos that have not yet been accreted onto larger structures. For comparison, our galaxy in figure  \ref{fig:ecc4trist90s} tumbles at a rate of $\sim 2 \pi / 5$ Gyrs$^{-1} \approx 1.7  $ h km s$^{-1}$ kpc$^{-1}$, which is an order of magnitude larger than their mean value. This is unsurprising, since, on average, the tidal torques experienced by substructure are much stronger than those that surround isolated halos.

\subsection{Intrinsic Rotation}

\begin{figure}
\begin{center}
{\includegraphics[width=9cm]{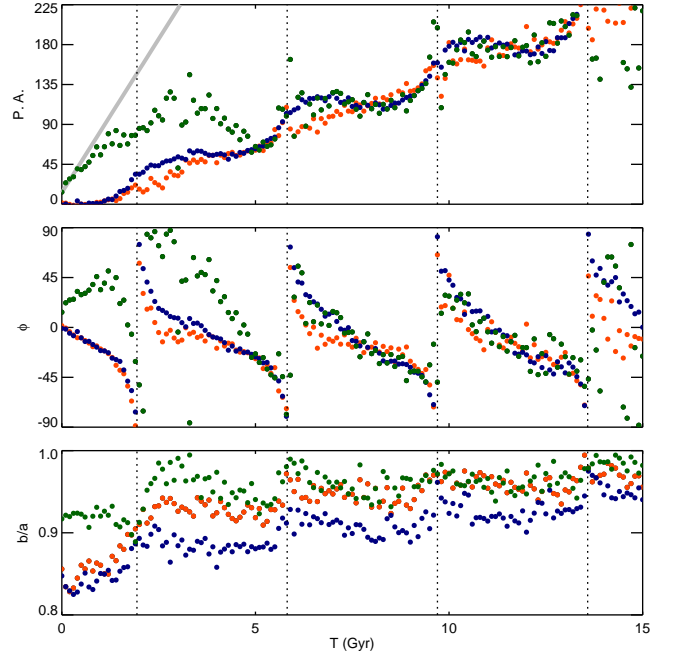}}
\caption[Triaxial galaxy with Initial Intrinsic Figure Rotation on an Eccentric Orbit]{\label{fig:ecc4trirot}\footnotesize{ Triaxial galaxy with intrinsic initial figure rotation, placed in eccentric (4:1) orbit around cluster potential. From top to bottom: Position angle, radial alignment and projected axis ratios of galaxy over 4 orbital periods. Gray line on top panel represents initial figure rotation of galaxy. The green points correspond to the rotating model, the red and blue are the non-rotating models from before. The dashed vertical lines represent pericenter passages}}
\end{center}
\end{figure}

If dark matter halos are already figure rotating slowly in isolation, will this make it harder for galaxies to become radially aligned once they fall inside clusters? A crude test of this was done by placing the galaxy described in the previous section, which has uniform figure rotation that is stable over many Gyrs, into the cluster again, but this time on the 4:1 eccentric orbit. The results of this experiment are shown in figure \ref{fig:ecc4trirot}. It is remarkable that the original rotation of the galaxy, represented by the thick gray line in the top panel, is quickly decelerated by the external tidal field of the cluster, and that, after 5 Gyrs, the behaviour of the initially rotating halo is essentially indistinguishable from the models with no rotation. In the first 5 Gyrs, the rotating galaxy is considerably less radially aligned than the non-rotating models, and, interestingly, this leads to a higher degree of sphericalization (see bottom panel), as predicted in \S \ref{sec:shapevol}. 

\section[Implications and Future Work]{Implications and Future Work}
\label{ch:imp}
\markright{}

\subsection{Dynamical Misalignments and Galaxy Modeling}
The experiments in this paper have shown that dynamical internal misalignments between dark matter and stellar light are an inevitable result of tidal torquing in external potentials. The precise radial twisting profile will depend on the galaxy's orbital phase and eccentricity, but generic models can be constructed from these  experiments that quantify the degree of misalignment vs radius, and this is the subject of a future paper. 

While some shape twisting is expected at larger radii ($r \ge r_{\rm t ,peri} \approx 50$ kpc) as a result of tidal stripping, the amount of twisting at small radii ($r \le r_{\rm s ,dm} \approx 22$ kpc)  is surprising, and will have important implications for mass modeling of galaxies in the cores of clusters. Strong lens models of clusters generally include bright galaxy members by adding pseudo-isothermal spherical mass distributions centered on the galaxies and truncated at some radius, typically $r_{\rm trunc} \approx 50$ kpc \citep{Natarajan:2009p970}. A few studies have started to use elliptical projected mass distributions for the dark matter, under the assumption that halos are triaxial and aligned with the luminous component of the galaxy. The amount of  shape twisting at $r< 50$ kpc seen in these simulations seems to invalidate this approach, and further work should be done to try and understand the impact of this twisting on strong lens models and cluster mass maps.

\subsection{Stacked Galaxy-Galaxy Lensing Measurements in Different Environments}

Quantifying the degree of internal misalignment in different galaxy populations is also vital for stacked galaxy-galaxy lensing techniques, where internal alignment is invariably assumed in statistical studies of dark matter halo ellipticities \citep[see, e.g., ][]{Mandelbaum:2006p1008}. One should be careful to exclude all galaxies in cluster environments from this type of study since this would dilute the measured ellipticity significantly.

On the other hand, it may be possible to make use of these misalignments to make an interesting measurement of dark matter substructure in clusters. \citet{Natarajan:2009p970} were able to detect tidal truncation of dark matter halos in the cluster Cl0024+16, by stacking ACS images of cluster galaxies in separate bins according to their distance from the cluster center. This type of analysis assumes galaxy dark matter halos align with the luminous galaxy within, and so images are rotated so that all optical major axes are coincident before stacking. However, if dark matter halos are in fact more strongly aligned toward the cluster center than their luminous counterparts, rotating these images so that they are all oriented in the direction of the cluster center before stacking will result in a stronger detection of ellipticity in the dark matter halo. This result would be an undeniable signature of dark matter that is impossible to explain within current theories of modified gravity.

\subsection{Weak Lensing Contamination}

Intrinsic alignments of galaxies are a serious contamination for large weak lensing and cosmic shear surveys, and one which has only recently started to receive focussed attention from the lensing community. It was until recently assumed that intrinsic alignments could only occur between galaxies at similar redshifts, which can then be easily downweighted in the cross-correlation analysis. Unfortunately the alignment mechanism described in this paper has the side effect of creating alignments between galaxies in widely separated reshift bins, since a galaxy that is radially aligned with the surrounding cluster potential will have an orientation that is anti-correlated with the orientations of the background galaxies being lensed by that same potential. This is a much harder contamination to account for, and requires a detailed understanding of galaxy alignments induced by tidal fields at all scales, not just in cluster potentials. Some attempts have been made recently to devise simple models that predict galaxy alignments according to the surrounding structure \citep[see, e.g.,][]{Schneider:2009p0}), but  further  work is required to fully incorporate the results and predictions of this paper into such a model. 

\subsection{Tidal Torquing of Disk Galaxies in Clusters}

Another possible consequence of the strong torquing of dark matter halos within hosts is disk warping. Because of their high angular momentum, disks will naturally resist tidal torquing more effectively than the surrounding dark matter halo, which will introduce a misalignment between the halo and disk. We have shown that tidal torquing affects all particles in the halos, even the most bound, so it is not unreasonable to expect that the inner shells should also feel these torques. 

There have been a few studies which attempt to address the issue of disk dynamics in tumbling dark matter halos \citep{Bureau:1999p2158,Bekki:2002pL21}. These studies find that a tumbling halo can induce significant warping in the disk, but they embed live disks in rigid, analytic halos with figure rotation, since it is difficult to generate N-body initial conditions for triaxial tumbling halos. This is probably not a a very realistic set-up, since, for a live halo, dynamical friction between the disk and halo will likely deform the halo potential, and bring the two into alignment. A recent paper by \cite{Dubinski:2009p2068} tries to overcome this issue by performing simulations on a self-consistent galaxy model with a live disk, bulge and halo component, which is then embedded in an analytic rotating potential which represents the tumbling outer halo. They find that this tumbling potential can induce significant disk warping and even trigger bar formation in a Milky Way like disk model. These studies were performed for isolated disk galaxies with dark matter halos that are tumbling at rates matching those measured in cosmological simulations. The work presented in this paper has shown that cluster potentials can also induce tumbling motion that is an order of magnitude larger in the halos of substructure galaxies. It will be interesting to see how disruptive this effect will be for a cluster disk galaxy, and this is work that is currently in progress.

\section{Conclusions}
\label{sec:conc}

The internal dynamics of galaxies in clusters are affected strongly by the cluster potential field at essentially all radii and certainly well below the traditional tidal radius.  It is a serious oversight to ignore tidal effects inside this radius when modeling the dynamical evolution of these galaxies. Galaxy orientations are affected by tides in two ways: in the outskirts, tidal fields distort galaxy shapes along the potential gradient over a short timescale. Closer in, tidal torquing of stellar orbits dominates, inducing figure rotation of the intrinsic triaxial shape.

In summary, our experiments show that: 

\begin{itemize}
\item On circular orbits, within an orbital period, galaxies are aligned radially and figure rotating at the same rate as their orbital motion, with no significant radial twisting of their shapes. The stellar component takes longer to become aligned, and this explains some of the discrepancy between the relatively weak observed signal, and the stronger alignment seen in the cosmological simulations of PBG08.

\item On eccentric orbits ($r_{\rm apo}/ r_{\rm peri} > 4$) the galaxy is mostly radially aligned throughout its orbit. The pericenter passage is too quick to change the stellar orientation significantly, affecting the galaxy mostly by heating its orbits, and driving the axis ratios to more spherical values. 

\item On intermediate orbits ($r_{\rm apo}/ r_{\rm peri} \sim 2 $), the galaxy's orientation is still controlled by external torques such that the mean orientation along the orbit remains radial. However, because the pericenter passage is slower and more disruptive, the stellar component is not as well aligned. The dark matter reacts more quickly, and is generally radially aligned throughout the orbit. 

\item The induced radial shape twisting within the galaxy leads to a misalignment between the stellar and dark matter orientations, since these have different radial extents. This contributes to the discrepancy between the observed strength of the dark matter and stellar alignment signals in previous works.

\item The figure rotation acquired by the orbiting galaxy becomes encoded in the stellar orbits, and is self sustaining for over 15 Gyrs in isolation.

\item Galaxies orbiting in cluster potentials tend to become more spherical with time: this is in part due to tidal shocking at pericenter, but it is also driven by tidal torquing during periods of radial misalignment.

\item Galaxies that fall into clusters with initial, intrinsic figure rotation will generally take longer to become radially aligned than non-rotating galaxies, although this rotation will be damped by the external torquing, and will eventually match the rotation acquired by the non-rotating models.
\end{itemize}

A particle's reaction to an external time varying tidal field depends strongly on its own orbital frequency. In the outer regions of a galaxy, the orbital frequencies of the dark matter particles are close to the frequencies of the satellite's own cluster orbit, which are in turn directly related to the variation in the external tide. In this regime, the orbital structure of particles is affected, leading to drastic changes in the shape and orientations of these outer layers on very short timescales. Stellar particles at the core of the galaxy orbit at much higher frequencies and thus see the cluster potential as changing more slowly, quasi-adiabatically. Their orbital structures are affected less, and their shapes are mostly preserved, although the orbital-averaged orientation still changes in response to an applied torque. 

This behavior leads to satellite shape profiles that can vary strongly as a function of radius. During pericenter passages, radial twists as large as \degrees{90} are observed within the dark matter scale radius, with correspondingly large misalignments between the stellar and dark matter orientations measured at different radii. These internal misalignments are also seen in the first few Gyrs of a circular orbit, before the satellite becomes tidally locked, as well as throughout most of the intermediate eccentricity orbits. 

We believe this internal twisting is the main reason behind the discrepancy between the radial alignment seen in cosmological dark matter simulations and the weaker alignment observed in the SDSS studies. While the degree of twisting in dark matter halos will be difficult to constrain observationally, these misalignments will strongly affect the mass modeling of cluster galaxies, and should be studied further.


\acknowledgements

We would like to thank Kathryn Johnston, Jeffrey Kenney, Jerry Sellwood and Jacqueline van Gorkom for their careful reading of an earlier version of this manuscript, and many helpful suggestions.  In addition, we acknowledge useful conversations with Christine Simpson.  M.J.P. acknowledges support from NASA grant NNX07AV06G; G.L.B. acknowledges support from NSF grants AST-05-07161 and AST-05-47823.  This research was supported in part by the NSF through TeraGrid resources provided by NCSA under grant MCA06T030, as well as by Columbia University through their support of the hotfoot cluster.

\bibliographystyle{apj}     

\bibliography{ms}

\end{document}